\documentclass[twocolumn,english,superscriptaddress,address,showpacs,showacknowledgments,onecolumn]{revtex4-1}
\usepackage[latin9]{inputenc}
\usepackage{babel}
\usepackage{amsmath}
\usepackage{amssymb}
\usepackage{graphicx}
\usepackage{esint}
\usepackage[unicode=true,pdfusetitle,
 bookmarks=true,bookmarksnumbered=false,bookmarksopen=false,
 breaklinks=false,pdfborder={0 0 1},backref=false,colorlinks=false]
 {hyperref}
\usepackage{breakurl}
\begin{document}

\title{Entanglement Rate for Gaussian Continuous Variable Beams}

\author{Zhi Jiao Deng}

\affiliation{Institut für Theoretische Physik II, Friedrich-Alexander-Universität
Erlangen-Nürnberg, Staudtstr. 7, D-91058 Erlangen, Germany}

\affiliation{Department of Physics, College of Science, National University of
Defense Technology, Changsha 410073, People's Republic of China}

\author{Steven J. M. Habraken}

\affiliation{Institut für Theoretische Physik II, Friedrich-Alexander-Universität
Erlangen-Nürnberg, Staudtstr. 7, D-91058 Erlangen, Germany}

\author{Florian Marquardt}

\email{Florian.Marquardt@physik.uni-erlangen.de}

\affiliation{Institut für Theoretische Physik II, Friedrich-Alexander-Universität
Erlangen-Nürnberg, Staudtstr. 7, D-91058 Erlangen, Germany}

\affiliation{Max Planck Institute for the Science of Light, Günther-Scharowsky-Straße
1/Bau 24, D-91058 Erlangen, Germany}

\pacs{03.67.Mn, 42.50.Dv, 42.50.Wk, 07.10.Cm }
\begin{abstract}
We derive a general expression that quantifies the total entanglement
production rate in continuous variable systems, where a source emits
two entangled Gaussian beams with arbitrary correlators. This expression
is especially useful for situations where the source emits an arbitrary
frequency spectrum, e.g. when cavities are involved. To exemplify
its meaning and potential, we apply it to a four-mode optomechanical
setup that enables the simultaneous up- and down-conversion of photons
from a drive laser into entangled photon pairs. This setup is efficient
in that both the drive and the optomechanical up- and down-conversion
can be fully resonant.
\end{abstract}
\maketitle

\section{Introduction}

Entanglement is an essential feature of quantum mechanics and a crucial
resource for  quantum communication and information processing. The
most common situation involves a source that continuously produces
entangled beams. One of the most natural characteristics of such a
source is obviously the rate at which it generates entanglement. If
the source sends out pairs of entangled particles, with subsequent
pairs completely independent, this rate can simply be defined as the
entanglement of each pair, divided by the time between pairs. However,
such a naive approach fails if there are correlations between subsequent
pairs, or if we consider entangled beams of radiation that cannot
be naturally decomposed into well-defined pairs of particles. In particular,
this is true for the very important case of continuous variable (CV)
entangled beams. Although many quantum information protocols exploit
qubits, with their discrete state space, the original Einstein-Podolsky-Rosen
\cite{EPR} entanglement involves continuous variables, and CV entanglement
has many modern applications \cite{DUAN,BOWEN,RCV,RMPCV,REID,ANDERSEN,SCHMIDT}.

In this article, we set out to provide a general definition for an
entanglement rate in such nontrivial situations. It will turn out
that our general definition, when applied to stationary Gaussian CV
beams, gives rise to a frequency integral over what we call a ``spectral
density of entanglement''. We show how to obtain this from the two-point
time correlators of the entangled beams, using a suitable additive
entanglement measure (the logarithmic negativity \cite{VIDAL}). For
one of the most common situations, we also provide explicit analytical
expressions. Our definition of the entanglement rate is particularly
important for setups where the output spectrum is arbitrary (e.g.
containing one or several peaks). This is a widespread case, since
the generation of CV-entangled radiation beams is often enhanced by
using cavity modes (e.g.~\cite{HUARD,WALLRAFF} in the microwave
domain, or \cite{KIMBLE,FUERST} in the optical domain). Recently,
the authors of the first experiment on spatially separated CV entanglement
in superconducting circuits even quantified their source by quoting
the effective number of entangled bits per second \cite{HUARD}, estimated
from the bandwidth of the circuit and the entanglement between two
modes. Our entanglement rate would provide a precisely defined way
to quantitatively characterize such situations. 

After we introduce and discuss the general definition, we illustrate
our entanglement rate by applying it to a four-mode optomechanical
setup that allows the fully resonant, and thereby efficient, generation
of entanglement. The rapidly developing field of cavity optomechanics
focuses on the dynamics of photons and phonons coupled via radiation
forces, see \cite{MARQUARDTREVIEW} for a recent review. The optomechanical
interaction has been predicted to produce entanglement, e.g. between
optics and mechanics \cite{BOOKCHAPTER,ZHANG,VITALI,AKRAM,HOFER,MARI-EISERT}
or between light modes \cite{PATERNOSTRO,MAVALVALA,GENES,VITALI-REVIEW,YIN,BARZANJEH2,BARZANJEH,TIAN,ROBUSTENTANGENK,WANG,Yingdan}.
Recently, optomechanical entanglement was demonstrated experimentally
for the first time, in a microwave circuit \cite{POLOMAKI}, making
its analysis especially timely.

\section{The entanglement rate}

\label{sec:EntanglementRate}

\begin{figure}
\includegraphics[width=1\columnwidth]{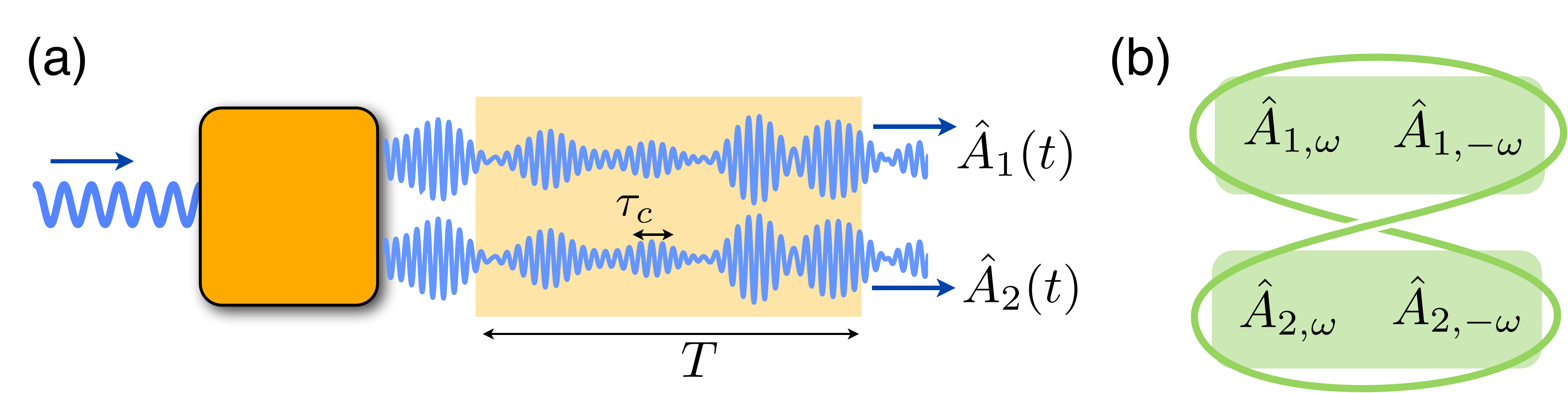}

\caption{\label{EntanglementRateFigure}(a) A source of two entangled beams,
where the aim is to calculate the entanglement between beams 1 and
2 in the large time interval $T$. (b) Frequency modes that are correlated
(entangled) in the limit of large $T$. }
\end{figure}

The situation we have in mind is very general: A source emits two
CV-entangled beams, described by bosonic fields $\hat{A}_{1}(t)$
and $\hat{A}_{2}(t)$, see Fig.~\ref{EntanglementRateFigure}a. These
could be, for example, two light beams of different polarization or
fields propagating along two different waveguides. Typical sources
might be a nonlinear crystal optical resonator driven by a pump beam,
a driven optomechanical cavity, or a driven on-chip microwave cavity
containing nonlinear elements like Josephson junctions. 

We focus on the important regime of generating stationary Gaussian
CV-entanglement. In that regime, one treats the pump as classical
and then obtains a quadratic Hamiltonian, leading to Gaussian statistics
of the emitted beams. Because of the pump, that Hamiltonian will be
time-dependent, containing terms of the form $\hat{a}_{c}^{\dagger}\hat{a}_{c}^{\dagger}e^{-i\omega_{p}t}+h.c.$,
where $\hat{a}_{c}$ would be a cavity mode (our analysis also applies
for several different modes). Here $\omega_{p}$ is the pump frequency
if the original nonlinearity was of the $\chi^{(2)}$-type, $\hat{a}_{c}^{\dagger}\hat{a}_{c}^{\dagger}\hat{a}_{p}+{\rm h.c.}$,
whereas $\omega_{p}$ would be twice the pump frequency for a $\chi^{(3)}$-type
nonlinearity $\hat{a}_{c}^{\dagger}\hat{a}_{c}^{\dagger}\hat{a}_{p}^{2}$.
As is usual in such situations, it will be most useful to switch to
a frame rotating at the frequency $\omega_{p}/2$, such that the Hamiltonian
becomes time-independent and we are dealing with a stationary problem.
In that new frame, $\omega=0$ relates to the pump frequency. 

Our analysis then focusses on the entanglement properties of the fields
$\hat{A}_{1,2}(t)$ emitted from any such source. Being Gaussian,
these fields are completely characterized by their two-time correlators.
The details of the source do not matter, except that it is assumed
to produce Gaussian beams that are stationary, i.e. where the correlators
only depend on the time-difference.

At this point we note that in some situations it may also be natural
to consider only a single field $\hat{A}(t)$, propagating along a
single waveguide. Then, frequency components centered symmetrically
around the pump frequency can be entangled, and they may afterwards
be directed to two different output ports by frequency-filtering.
For such a situation, we can still apply our approach if we define
$\hat{A}_{1}(t)$ to contain the positive frequency ($\omega>0$)
components of the field $\hat{A}(t)$, while likewise $\hat{A}_{2}(t)$
would contain the negative frequency ($\omega<0$) components, where
$\omega$ is already determined with respect to the rotating frame.

Since the situation is stationary, it is natural to try and define
an \emph{entanglement} \emph{rate}, i.e. the entanglement per unit
time emitted from the source. We propose to do this in the most natural
way by calculating the \emph{overall} amount of entanglement between
the two beams in a long time interval $T$ and then dividing by $T$,
in the end sending $T$ to infinity:

\begin{equation}
\Gamma_{E}\equiv\lim_{T\rightarrow\infty}\frac{E(T)}{T}\,.
\end{equation}
This definition is not constrained to CV entangled beams or to Gaussian
states. It only requires (a) stationarity of the source, (b) an entanglement
measure that is additive for product states, and (c) a finite correlation
time $\tau_{c}$ for the beams. Given such a finite correlation time,
the fields on two subsequent time-intervals of length $T\gg\tau_{c}$
are not correlated to a very good approximation. This holds because
even though there may be remaining correlations near the boundary
between the time-intervals, these are appreciable only up to a distance
of order $\tau_{c}$ from the boundary and can be neglected in the
limit $T\gg\tau_{c}$. Due to the additivity (and stationarity), we
then get $E(2T)\approx2E(T)$, such that the entanglement rate calculated
for time intervals of sizes $T$ or $2T$ is the same (up to the small
corrections which we can neglect in the limit of large $T$).

From now on, we specialize to stationary Gaussian CV entangled beams.
We will use the logarithmic negativity \cite{VIDAL,RCV} $E_{N}$
as an entanglement measure, since it is both straightforwardly evaluated
for Gaussian multi-mode states and has the important property of additivity.

Let us now consider the fields $\hat{A}_{s}(t)$ on the interval $[0,T]$,
even though they are defined for all times $t$. We can define discrete
frequency modes ($s=1,2$):

\begin{equation}
\hat{A}_{s}(t)=\frac{1}{\sqrt{T}}\sum_{\omega=j2\pi/T}\hat{A}_{s,\omega}e^{-i\omega t}\,.
\end{equation}
The normalization is chosen such that the $\hat{A}_{s,\omega}$ fulfill
bosonic commutation relations, $\left[\hat{A}_{s,\omega},\hat{A}_{s',\omega'}^{\dagger}\right]=\delta_{s,s'}\delta_{\omega,\omega'}$,
where $\omega=j2\pi/T$ is discrete, with $j$ an integer. 

We now want to calculate the full logarithmic negativity $E_{N}(T)$
between the two beams on that time interval, which is equivalent to
the entanglement between two sets of harmonic oscillators \cite{Eisert}.
In our case, we are even considering infinitely many harmonic oscillators
$\hat{A}_{s,\omega}$. We stress that the entanglement $E_{N}(T)$
is (of course) independent of our choice of basis for each of the
beams, as a different choice of basis amounts to implementing a local
unitary transformation. The correlations between the two beams can
be arbitrary, except that they are supposed to decay beyond some correlation
time $\tau_{c}$ (which is true in any reasonable physical situation).
As already mentioned above, we will assume that $T\gg\tau_{c}$, such
that the correlations between subsequent intervals of size $T$ can
be neglected. In that limit, we can use stationarity to show that
only the following types of correlators may be nonzero in the present
situation, up to corrections that are small in $\tau_{c}/T$: $\left\langle \hat{A}_{s,\omega}^{\dagger}\hat{A}_{s',\omega}\right\rangle $
and $\left\langle \hat{A}_{s,\omega}\hat{A}_{s',-\omega}\right\rangle $
(and their conjugates). For example, we find

\begin{equation}
\left\langle \hat{A}_{s,\omega}^{\dagger}\hat{A}_{s',\omega'}\right\rangle \approx\delta_{\omega,\omega'}\left\langle \hat{A}_{s}^{\dagger}\hat{A}_{s'}\right\rangle _{-\omega}\,,\label{eq:Correlator}
\end{equation}
where we have defined the Fourier transform of the correlator: $\left\langle \hat{A}\hat{B}\right\rangle _{\omega}\equiv\int_{-\infty}^{+\infty}dt\,e^{i\omega t}\left\langle \hat{A}(t)\hat{B}(0)\right\rangle $.
Likewise, we have $\left\langle \hat{A}_{s,\omega}\hat{A}_{s',\omega'}\right\rangle =\delta_{\omega,-\omega'}\left\langle \hat{A}_{s}\hat{A}_{s'}\right\rangle _{\omega}$. 

In summary, for any given $\omega>0$, only the correlators involving
the modes $\omega$ and $-\omega$ are nonzero (see Fig.~\ref{EntanglementRateFigure}b).
There is entanglement $E_{N}[\omega]$ between the two modes $\hat{A}_{1,\omega}$
and $\hat{A}_{1,-\omega}$ of beam 1 with the two modes $\hat{A}_{2,\omega}$
and $\hat{A}_{2,-\omega}$ of beam 2. Further below, we will show
how $E_{N}[\omega]$ can be calculated from the correlators. The overall
entanglement $E_{N}(T)$ between the two beams can then be decomposed
into a sum, because of the additivity of the logarithmic negativity:

\begin{equation}
E_{N}(T)=\sum_{\omega>0}E_{N}[\omega]\,.
\end{equation}
We note that $E_{N}[\omega]$ in the sum does not depend on $T$ (after
adopting the approximation of Eq.~\ref{eq:Correlator}), although
the number of summation terms does scale with $T$ due to the discretization
of $\omega$. 

The sum over discrete frequencies can be converted into an integral.
To leading order at large $T$, we have $\sum_{\omega>0}E_{N}[\omega]\approx(2\pi/T)^{-1}\int_{0}^{\infty}d\omega\,E_{N}[\omega]$.
Therefore, we can identify $E_{N}[\omega]$ as the \emph{``spectral
density of entanglement''} \cite{FOOTNOTEENTANGLEMENT}. In addition,
this confirms that the total entanglement indeed grows linearly in
$T$ for sufficiently large times. As a consequence we finally obtain
the \emph{entanglement rate}:

\begin{equation}
\Gamma_{E}\equiv\lim_{T\rightarrow\infty}\frac{E_{N}(T)}{T}=\int_{0}^{\infty}\frac{d\omega}{2\pi}E_{N}[\omega]\,.
\end{equation}
This shows that $E_{N}[\omega]$ itself may be interpreted as the
\emph{entanglement rate per frequency interval}. It is possible to
give an explicit expression for $E_{N}[\omega]$ in terms of the covariance
matrix containing the correlators \cite{VIDAL}. The general case,
involving the four modes $\hat{A}_{1,\omega}$, $\hat{A}_{1,-\omega}$,
$\hat{A}_{2,\omega}$, and $\hat{A}_{2,-\omega}$, is a bit cumbersome,
requiring the symplectic diagonalization of a $8\times8$ covariance
matrix. However, the expressions simplify considerably if there is
purely two-mode squeezing, i.e. if the intra-mode correlators $\left\langle \hat{A}_{s,\omega}\hat{A}_{s,-\omega}\right\rangle $
vanish. In that case, we find

\begin{equation}
E_{N}[\omega]=E[\omega]+E[-\omega]\,.
\end{equation}
For positive $\omega$, $E[\omega]$ is the entanglement between $\hat{A}_{1,\omega}$
and $\hat{A}_{2,-\omega}$ while $E[-\omega]$ is the entanglement
between $\hat{A}_{1,-\omega}$ and $\hat{A}_{2,\omega}$. In contrast
to $E_{N}[\omega]$, the density $E[\omega]$ is double-sided (has
contributions both at negative and positive frequencies). We note
that in the following we will also commonly refer to $E[\omega]$
as the spectral density of entanglement, since it is closely related
to $E_{N}[\omega]$. Setting $n_{+}\equiv\left\langle \hat{A}_{1,\omega}^{\dagger}\hat{A}_{1,\omega}\right\rangle +\frac{1}{2}$,
$n_{-}\equiv\left\langle \hat{A}_{2,-\omega}^{\dagger}\hat{A}_{2,-\omega}\right\rangle +\frac{1}{2}$,
and $\xi\equiv\left\langle \hat{A}_{1,\omega}\hat{A}_{2,-\omega}\right\rangle $,
we have:

\begin{eqnarray}
E[\omega] & = & {\rm max}(0,-\ln(2\eta_{-})),\label{eq:EomegaExplicit}\\
2\eta_{-} & = & n_{+}+n_{-}-\sqrt{(n_{+}-n_{-})^{2}+4\left|\xi\right|^{2}}\,.
\end{eqnarray}
As an aside we note that we choose to work with the natural logarithm
in our discussion (some articles use $\log_{2}$, which is more natural
for discrete qubits \cite{VIDAL}). 

In this special case, the entanglement rate is therefore:

\begin{equation}
\Gamma_{E}=\int_{-\infty}^{+\infty}\frac{d\omega}{2\pi}E[\omega]\,.\label{eq:GammeEomega}
\end{equation}

\section{Relation to Entanglement between Wave Packets}

We now want to connect our general result to previously applied approaches
for quantifying the entanglement in such situations. It is a common
procedure to employ normalized mode functions (``filter functions'')
$f_{s}(t)$ that have the shape of wave packets, in order to define
two harmonic oscillator modes, one for each beam:

\begin{equation}
\hat{a}_{s}\equiv\int_{-\infty}^{+\infty}dt\,f_{s}(t)\hat{A}_{s}(t)\,.
\end{equation}
Here $\int_{-\infty}^{+\infty}\left|f_{s}(t)\right|^{2}dt=1$. The
entanglement between $\hat{a}_{1}$ and $\hat{a}_{2}$ can then be
calculated, e.g. again using the logarithmic negativity as an entanglement
measure. This filtering analysis has been applied to several settings
\cite{GENES,BARZANJEH2,ROBUSTENTANGENK}, for a detailed explanation
see e.g. \cite{GENES,VITALI-REVIEW}. The catch is that this procedure
introduces a filtering time $\tau$ (the extent of the wave packets),
and the results will be a function of $\tau$, which is usually taken
to be arbitrary.

How would one loosely define an entanglement rate based on this procedure?
We can imagine that there is a stream of such wave packets, with a
spacing of about $\tau$ (where care would have to be taken to define
them to be orthogonal). A simple, though phenomenological approach
to define an entanglement rate would be to simply calculate the ratio
$E_{N}^{\tau}/\tau$. 

However, it is clear that this approach is not systematic. In fact,
it cannot always cover the full entanglement, since there may be entanglement
in components of the beam that are orthogonal to the filter functions
which are employed. In addition, there is some arbitrariness in the
choice of filter function (and, thus, even in the definition of $\tau$).
Moreover, one may have situations where there are temporal correlations
extending beyond $\tau$. Then, the entanglement present in the beams
may be underestimated. If one tries to remedy this problem by choosing
larger $\tau$, then the filter bandwidth shrinks and one may miss
entanglement present at other frequencies.

It turns out that an approach based on wave packets can be made to
work, but only if one constructs a suitable complete basis that has
a clear physical meaning. It is more systematic than the naive filtering
approach described so far and it covers all the entanglement. In addition,
it can be related to a direct physical prescription, and we will see
that it leads to the same results as our general, basis-independent
definition discussed in the previous section.

\begin{figure}
\includegraphics[width=1\columnwidth]{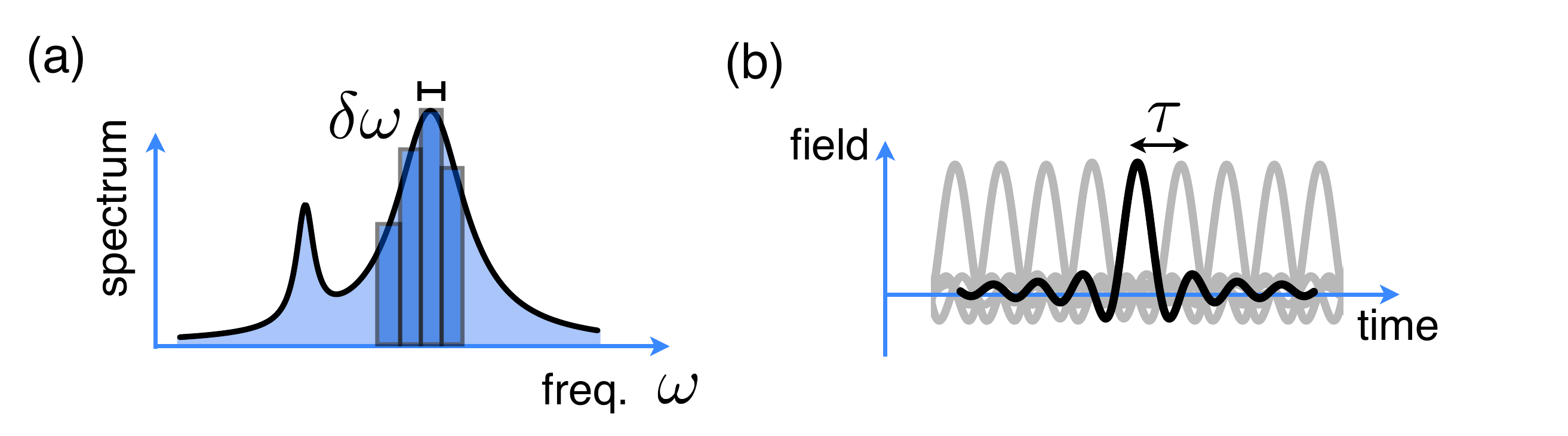}

\caption{\label{Fig2}Wave packet picture that can be used in understanding
the entanglement rate. (a) For each of the beams, we imagine to frequency-filter
the field, with a filter size $\delta\omega$. (b) In time-space,
this corresponds to wave packets spaced by $\tau=2\pi/\delta\omega$.
These form a complete basis on a time-frequency grid. The use of a
wave packet basis on a time-frequency grid in the context of quantum
noise is reviewd in \cite{asha}.}

\end{figure}

For simplicity, focus on the situation with only cross-correlations
(no intra-beam squeezing) that we discussed at the end of section
\ref{sec:EntanglementRate}. Imagine one sends one beam through a
frequency filter $[\omega,\omega+\delta\omega]$, where $\delta\omega=2\pi/\tau$.
Likewise, the other beam will be sent through another filter, at negative
frequencies $[-\omega,-\omega-\delta\omega]$. Now construct a complete
set of orthogonal wave packet modes (``Wannier basis'') with a spacing
$\tau$ in time, which are able to fully represent the filtered beams
(see Fig.~\ref{Fig2}). As we show in the appendix \ref{AppendixWavePacketBasisGeneral},
the logarithmic negativity $E_{N}^{\tau}$ between two such wave packet
modes (one in each beam, at equal time-slots) is related to $E[\omega]$
in the limit $\tau\gg\tau_{c}$:

\begin{equation}
\lim_{\tau\rightarrow\infty}E_{N}^{\tau}=E[\omega]\,.
\end{equation}

As a consequence, the general definition of the previous section agrees
with the entanglement rate calculated from such a wave packet picture.
This wave packet approach can also be viewed as representing the following
physical procedure: Split each of the two entangled beams into many
frequency-filtered output beams, where the frequency resolution $\delta\omega=2\pi/\tau$
has been chosen fine enough, such that $\delta\omega\ll1/\tau_{c}$.
The rate $\Gamma_{E}$ quantifies the total entanglement per unit
time contained in the sum of those streams (since it was defined that
way in the previous section). Each pair of wave packets (of length
$\tau$) can in principle be exploited for an application such as
CV quantum teleportation. A concrete physical measurement of the entanglement
between any two wave packets could be performed in a standard way,
using homodyne measurements. For example, the local oscillator can
provide strong pulses that are shaped in the form of the wave packet
modes that we want to consider.

\section{A specific optomechanical example, and its implementation}

\subsection{Model}

We illustrate the features of the entanglement rate in a model describing
the effective interaction between two localized optical modes ($\hat{a}_{+}$
and $\hat{a}_{-}$) and a mechanical mode ($\hat{b}$):

\begin{eqnarray}
\hat{H}/\hbar=-\Delta(\hat{a}_{+}^{\dagger}\hat{a}_{+}+\hat{a}_{-}^{\dagger}\hat{a}_{-})+\delta\hat{b}^{\dagger}\hat{b}\nonumber \\
-\frac{g}{2}\{(\hat{a}_{+}+\hat{a}_{-}^{\dagger})\hat{b}^{\dagger}+(a_{+}^{\dagger}+\hat{a}_{-})\hat{b}\}\;,\label{Hlin}
\end{eqnarray}
Similar Hamiltonians have been studied previously in the context of
entanglement generation between light modes \cite{GENES,PATERNOSTRO,BARZANJEH,BARZANJEH2,YIN,TIAN,ROBUSTENTANGENK,WANG,MAVALVALA,VITALI-REVIEW,Yingdan}.
For instance, Wang and Clerk \cite{WANG} studied intra-cavity entanglement,
while Tian \cite{TIAN} investigated also the stationary output entanglement.
We note that, more recently, an analysis of the output entanglement
in the setup with $\Delta=\delta=0$ (but with unequal couplings for
up- and down-conversion) was provided in \cite{Yingdan}.

\begin{figure}[t]
\includegraphics[width=1\columnwidth]{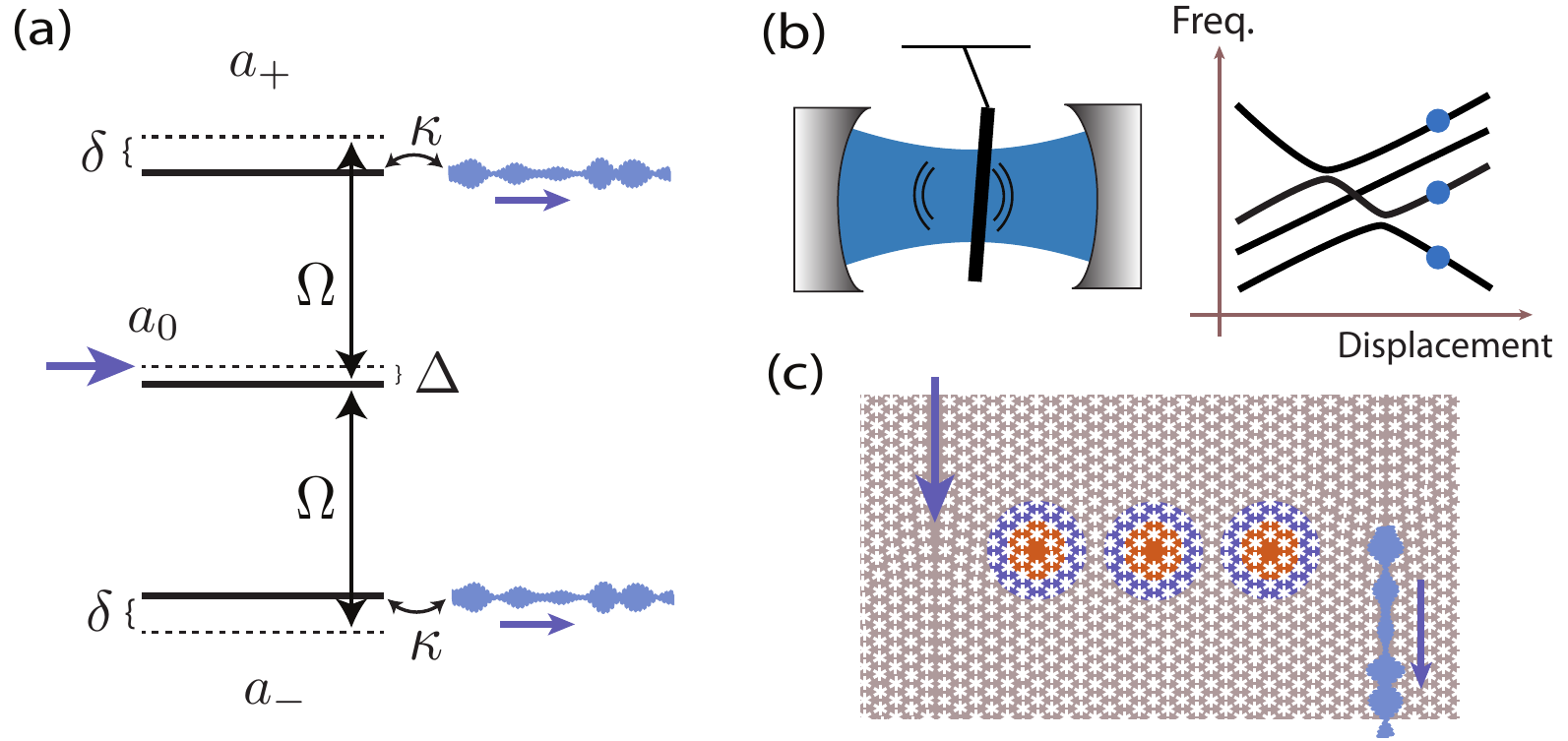} \caption{\label{SetupFigure}(a) Level diagram: the optical level distance
almost matches the vibrational frequency $\Omega$ so that phonon-mediated
transitions between the optical levels occur. (b) Schematic for a
possible implementation in a membrane-in-the-middle setup; three of
the optical modes constitute an equidistant triplet for a suitable
membrane position and tilt angle. (c) Schematic for the possible implementation
in an optomechanical crystal, with one incoupling and one outcoupling
waveguide.}
\end{figure}

Before proceeding, we note how to implement this model  using three
equidistant optical modes, enhancing the efficiency beyond previous
suggestions. The optical mode spacing $J$ is nearly resonant with
the vibration frequency $\Omega$, with a frequency mismatch $\delta=\Omega-J$
(Fig.~\ref{SetupFigure}a). A laser drives the center optical mode
at $\omega_{0}$, with a detuning $\Delta=\omega_{L}-\omega_{0}$.
An optomechanical interaction of the kind $\hbar g_{0}\hat{a}_{+}^{\dagger}\hat{a}_{0}(\hat{b}+\hat{b}^{\dagger})$,
and similarly for $\hat{a}_{-}$, scatters photons up and down, into
modes $\hat{a}_{+}$ and $\hat{a}_{-}$, while simultaneously destroying
(creating) phonons. When a phonon is virtually emitted and re-absorbed,
an effective four-wave mixing process is induced, generating a pair
of $\hat{a}_{+}$ and $\hat{a}_{-}$ photons out of a pair of $\hat{a}_{0}$
photons. Thus, two-mode vacuum squeezing (EPR entanglement) is produced.
We assume that the drive is strong and we can replace $\hat{a}_{0}$
by the coherent amplitude $\alpha=\left\langle \hat{a}_{0}\right\rangle $.
This yields the Hamiltonian (\ref{Hlin}) with $g\equiv2g_{0}\alpha$,
provided we choose frames rotating at $\omega_{L}\pm J$ for the modes
$\hat{a}_{\pm}$, and at $J$ for $\hat{b}$. Moreover, only nearly
resonant terms are kept, which is allowed if $J,\,\Omega\gg g,\,\kappa$,
where $\kappa$ is the optical intensity decay rate.

Possible experimental implementations include a membrane-in-the-middle
setup tuned to a point with three equidistant modes \cite{HARRIS2008,HARRIS2,HARRIS3}
(Fig.~\ref{SetupFigure}b) or coupled optomechanical cells, e.g.
in an optomechanical crystal \cite{OMCRYSTALS,SNOWFLAKE} (Fig.~\ref{SetupFigure}c),
see the appendix. Such a triply-resonant setup enhances the efficiency:
As compared to previous suggestions with only one resonantly driven
mode, generating entangled Stokes and anti-Stokes sidebands (similar
to \cite{GENES}), one wins a factor $\Omega^{4}/(\delta^{2}+(\kappa/2)^{2})^{2}\gg1$
in the intensity of the entangled output beams, while compared to
setups with two optical ouput modes (e.g. \cite{TIAN,WANG}), one
wins a factor $(2\Omega/\kappa)^{4}\gg1$, for fixed input laser power
and $\Delta=0$.

We use standard input-output theory\cite{INPUTOUTPUT} for our analysis:

\begin{eqnarray}
\dot{\hat{a}}_{j}=\frac{i}{\hbar}[\hat{H},\hat{a}_{j}]-\frac{\kappa}{2}\hat{a}_{j}-\sqrt{\kappa}\hat{a}_{j,{\rm in}}(t),\nonumber \\
\dot{\hat{b}}=\frac{i}{\hbar}[\hat{H},\hat{b}]-\frac{\Gamma}{2}\hat{b}-\sqrt{\Gamma}\hat{b}_{{\rm in}}(t)\label{InputOutput}
\end{eqnarray}
Here $\Gamma$ is the mechanical damping rate and $j=\pm$. As usual,
$\langle\hat{b}_{{\rm in}}^{\dagger}(t)\hat{b}_{{\rm in}}(0)\rangle=n_{{\rm th}}\delta(t)$
and $\langle\hat{b}_{{\rm in}}(t)\hat{b}_{{\rm in}}^{\dagger}(0)\rangle=(n_{{\rm th}}+1)\delta(t)$,
with $n_{{\rm th}}=(\exp(\hbar\Omega/(k_{{\rm B}}T))-1)^{-1}$ the
thermal occupation, and likewise for $\hat{a}_{j,{\rm in}}$ (but
without thermal noise). Solving Eqs.~(\ref{InputOutput}) and employing
$\hat{a}_{j,{\rm out}}(t)=\hat{a}_{j,{\rm in}}(t)+\sqrt{\kappa}\hat{a}_{j}(t)$,
we find the linear relation between output and input fields in terms
of a scattering matrix.

Coming back to our general definition of the entanglement rate, we
would consider $\hat{a}_{+,{\rm out}}(t)$ as the first beam $\hat{A}_{1}(t)$
and $\hat{a}_{-,{\rm out}}(t)$ as the second beam $\hat{A}_{2}(t)$.
It is not difficult to show (and can be confirmed by direct calculation)
that there is no intra-beam squeezing, i.e. $\left\langle \hat{A}_{1,\omega}\hat{A}_{1,-\omega}\right\rangle =0$
and likewise for beam 2. Thus, we want to employ the formulas Eqs.
(\ref{eq:EomegaExplicit}) and (\ref{eq:GammeEomega}) in order to
find the contributions to the spectral density of entanglement and
the entanglement rate. 

To calculate $E[\omega]$, we need the correlators of the two beams.
Entangled photon pairs are emitted at \emph{physical} frequencies
$\omega_{L}\pm J\pm\omega$, corresponding to $\pm\omega$ in our
rotating frame. We have to evaluate correlators like $\int_{-\infty}^{+\infty}e^{i\omega t}\left\langle \hat{a}_{+,{\rm out}}(t)\hat{a}_{-,{\rm out}}(0)\right\rangle \,dt$
as shown in appendix \ref{AppendixCalcEntanglement}.

\begin{figure}[t]
\includegraphics[width=1\columnwidth]{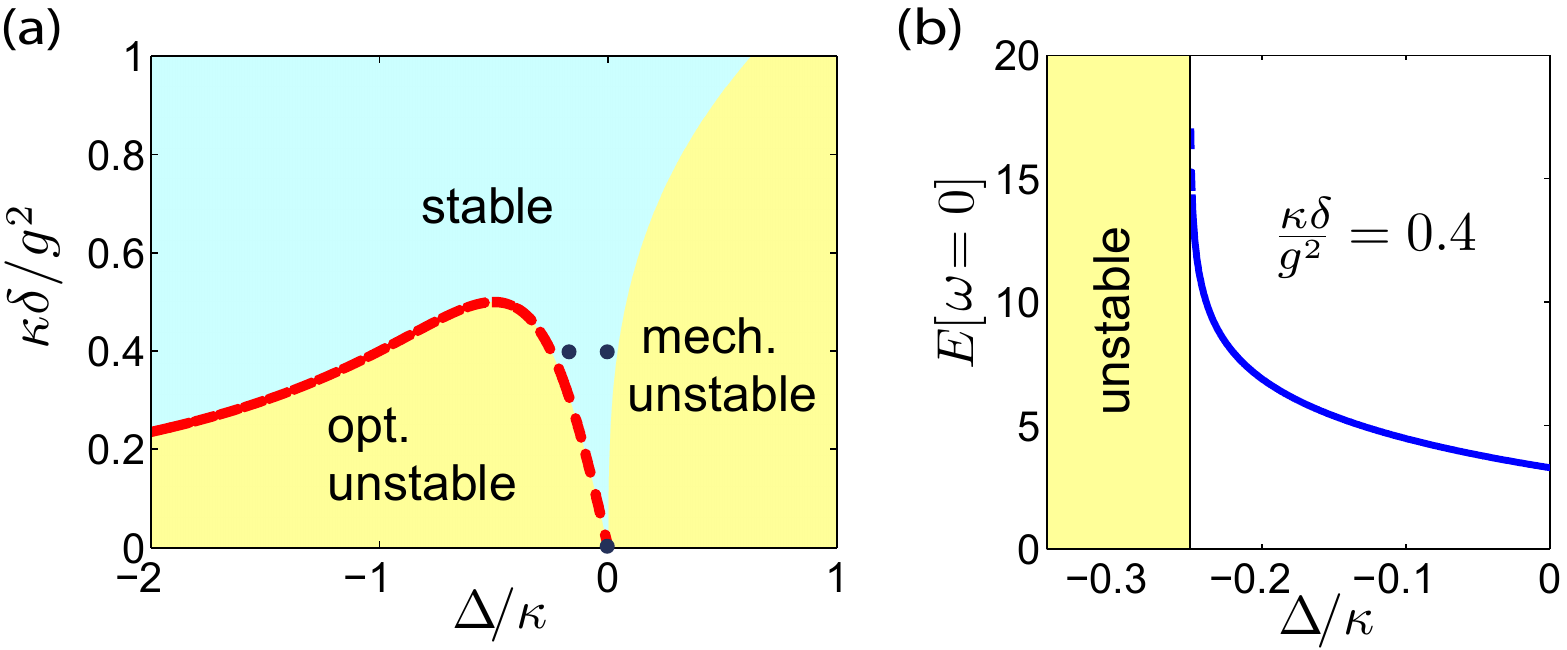} \caption{\label{StabilityDiagram}(a) Stability diagram of the model, vs. frequency
mismatch $\delta$ and laser detuning $\Delta$, for $g/\kappa=5$
and $\Gamma/\kappa=10^{-3}$. Stability boundary in the effective
optical model: red dashed line. Parameter values corresponding to
the (stable) blue points will be studied below. (b) Diverging entanglement
from the effective optical model at the boundary of stability.}
\end{figure}

\subsection{Results}

We first address some general features. The system can become unstable
(Fig.~\ref{StabilityDiagram}a), both towards mechanical and optical
oscillations. The optical stability boundary is approximately given
by $(\Delta+g^{2}/2\delta)\Delta+(\kappa/2)^{2}=0$. Eliminating the
mechanical mode (for $\delta\gg\kappa,\,\Gamma,\,g,\,|\Delta|$),
we obtain an effective optical model, which yields the photon pair
creation rate $\Gamma_{{\rm pairs}=}\left(\frac{g^{2}}{4\delta}\right)^{2}\frac{\kappa}{2}[(\kappa/2)^{2}+(\triangle+\frac{g^{2}}{2\delta})\triangle]^{-1}$.
This diverges at the optical stability boundary.

We now focus on the spectral density of entanglement $E[\omega]$
that characterizes the output beams, and especially the entanglement
rate. In contrast to the intra-cavity entanglement (discussed in \cite{WANG,TIAN}),
we find that $E[\omega]$ is not bounded. This is similar to the difference
between intra-cavity and output squeezing \cite{COLLET,GEA,MILBURN}.
Numerical plots of $E[\omega]$ for the Hamiltonian (\ref{Hlin})
have been shown so far \cite{TIAN} only for the special case $\Delta=\delta=0$
and asymmetric optomechanical couplings. Entanglement of temporal
modes was also discussed for a pulsed scheme \cite{ROBUSTENTANGENK}
in the case $\delta\neq0$. 

The output entanglement grows significantly at the stability boundary
(Fig.~\ref{StabilityDiagram}b), even diverging in the effective
optical model. This is typical near an instability but it comes at
the price of linewidth narrowing, reducing the entanglement rate.

\begin{figure}
\includegraphics[width=1\columnwidth]{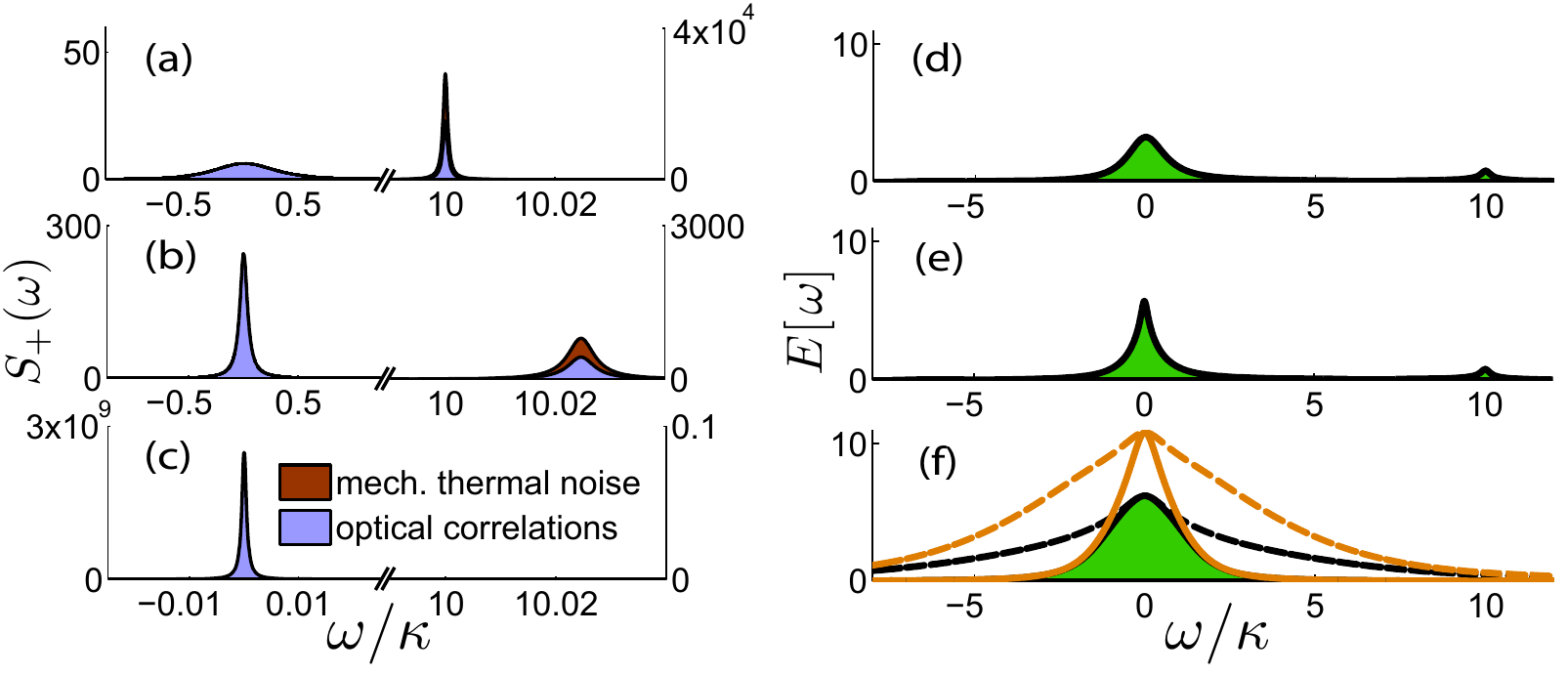}

\caption{\label{OutputSpectrum}The output spectrum $S_{+}(\omega)$ for the
beam from the upper optical mode, plotted for three typical situations
(a,b,c) at a fixed temperature ($n_{{\rm th}}=50$), with $g/\kappa=5$
and $\Gamma/\kappa=10^{-3}$. Blue vs. brown distinguish optical vs.
mechanical noise contributions. (a) resonant drive, off-resonant mechanical
mode ($\Delta=0,\,\delta=10\kappa$); (b) near the optical instability
($\Delta=-0.2\kappa,\,\delta=10\kappa$); (c) doubly resonant ($\Delta=\delta=0$).
Note the differences in vertical and horizontal scales for the peaks.
(d,e,f) The spectral density of entanglement $E[\omega]$, corresponding
to (a,b,c). Additional curves in (f) correspond to $\Gamma/\kappa=5\cdot10^{-2}$(dashed)
and $\Gamma/\kappa=10^{-3}$ (solid), for $n_{{\rm th}}=0$ (orange)
and $n_{{\rm th}}=50$ (black), at fixed cooperativity $\mathcal{C}=g^{2}/\kappa\Gamma=2.5\cdot10^{4}$.}
\end{figure}

In order to appreciate this, we now discuss the output intensity spectrum,
$S_{+}(\omega)=\int e^{-i\omega t}\left\langle \hat{a}_{+,{\rm out}}^{\dagger}(t)\hat{a}_{+,{\rm out}}(0)\right\rangle dt$,
in Fig.~\ref{OutputSpectrum}a,b,c. We expect that any mechanical
noise contributing to the optical output is deleterious for entanglement,
which is confirmed by explicit calculation. Thus, we plot the spectrum
in an instructive way, distinguishing optical and mechanical contributions,
as obtained from the linear relation $\hat{a}_{+,{\rm out}}(\omega)=\ldots\hat{a}_{+,{\rm in}}(\omega)+\ldots(\hat{a}_{-,{\rm in}}^{\dagger})(\omega)+\ldots\hat{b}_{{\rm in}}(\omega)$.
There are typically two peaks, separated by $\delta=\Omega-J$ and
containing primarily optical or mechanical noise, respectively. Near
the optical instability (Fig.~\ref{OutputSpectrum}b), the optical
peak gets strong and narrow. 

When the optical mode spacing matches the mechanical frequency ($\delta=0$)
and the laser is on resonance ($\Delta=0$), the two peaks merge and
have a narrow linewidth set by the mechanical damping rate $\Gamma$
( Fig.~\ref{OutputSpectrum}c). It will turn out that the entanglement
rate is maximized near this point. This is entirely counterintuitive:
One might expect that at $\delta=0$ mechanical noise is injected
into the output beams, destroying entanglement. However, we find that
the optical noise can completely overwhelm the mechanical noise for
strong driving, when the cooperativity is sufficiently large, $\mathcal{C}\equiv g^{2}/\kappa\Gamma\gg n_{{\rm th}}$. 

The spectral density $E[\omega]$ is shown in Fig.~\ref{OutputSpectrum}d,e,f.
Typically, $E[\omega]$ is maximal at the optical peak near $\omega=0$.
For the important case $\Delta=\delta=0$, we find an analytical expression,
$E[\omega=0]=-\ln(2\eta_{-})$, where

\begin{equation}
\eta_{-}=4\mathcal{C}(\mathcal{C}+n_{{\rm th}}+\frac{1}{2})+\frac{1}{2}-2\mathcal{C}\sqrt{1+4(\mathcal{C}+n_{{\rm th}}+\frac{1}{2})^{2}}
\end{equation}
depends on both driving (via $\mathcal{C}$) and temperature. We checked
that choosing different optomechanical couplings for the two modes
$\hat{a}_{\pm}$ will not increase $E[\omega=0]$, in contrast to
the intra-cavity case \cite{WANG}. 
\begin{figure}[t]
\includegraphics[width=1\linewidth]{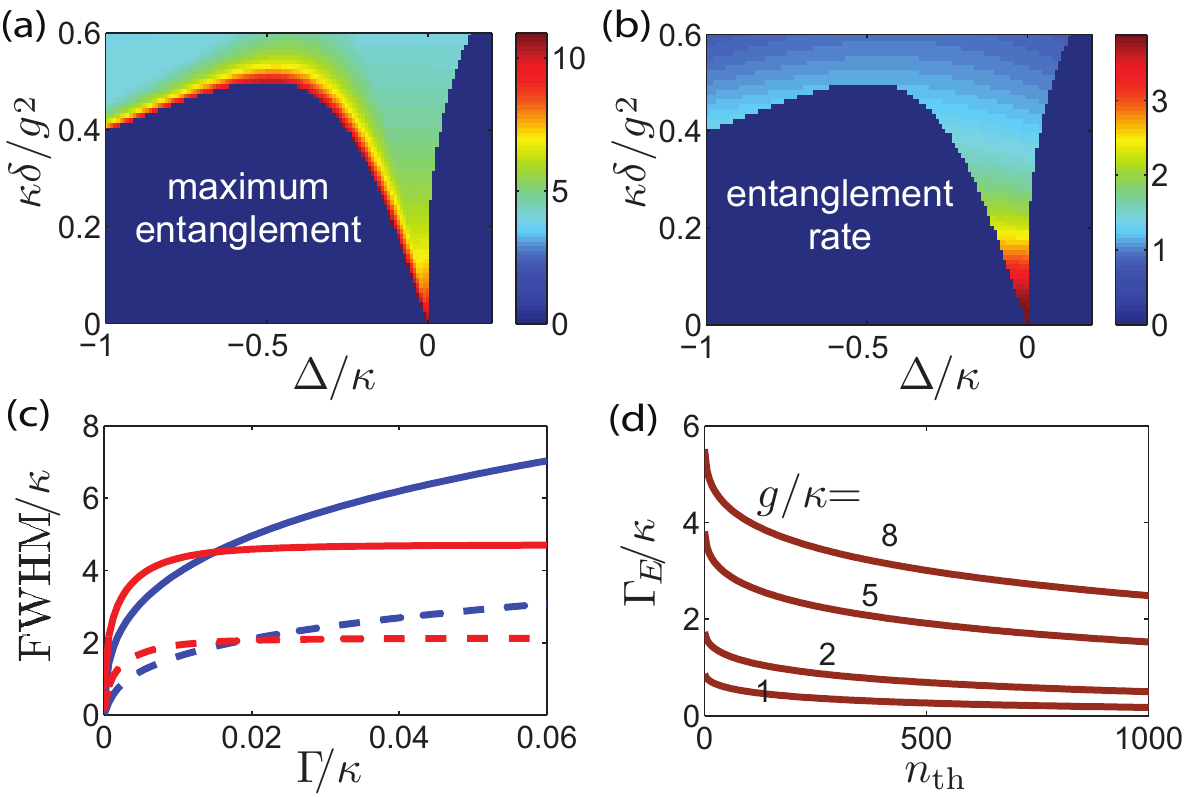} \caption{\label{EntanglementComparison}(a) Spectral density of entanglement,
maximized over frequency, ${\rm max}_{\omega}E[\omega]$, and (b)
total entanglement rate $\Gamma_{E}=\int E[\omega]d\omega/2\pi$ (in
units of $\kappa$), both shown as a function of frequency mismatch
$\delta$ and laser detuning $\Delta$, for $n_{{\rm th}}=0$, $g/\kappa=5$
and $\Gamma/\kappa=10^{-3}$. (c) The full-width-half-maximum (FWHM)
of the peak in $E[\omega]$ as a function of $\Gamma/\kappa$, for
fixed cooperativity $\mathcal{C}=2.5\cdot10^{4}$ (solid) and $\mathcal{C}=10^{3}$
(dashed), at $\delta=\Delta=0$. (Blue: $n_{{\rm th}}=0$, Red : $n_{{\rm th}}=50$).
(d) Temperature dependence of $\Gamma_{E}$ ($\delta=\Delta=0$ and
$\Gamma/\kappa=10^{-3}$).}
\end{figure}

In Fig.~\ref{EntanglementComparison}a,b, we compare the maximum
of the spectral density of entanglement, $E_{{\rm max}}\equiv{\rm max}_{\omega}E[\omega]$,
and the entanglement rate $\Gamma_{E}=\int E[\omega]d\omega/2\pi$.
While $E_{{\rm max}}$ becomes large near the optical boundary of
stability, the entanglement rate there remains small, due to the narrow
bandwidth. This will be a general feature in many similar systems.
Rather, $\Gamma_{{\rm E}}$ is optimal near $\delta=\Delta=0$. We
found that, as the mechanical damping rate increases, the optimum
shifts away from $\Delta=\delta=0$. 

The entanglement rate depends on the full shape of $E[\omega]$, in
particular the peak width(s). Crucially, those widths are distinct
from those in the output spectrum, due to the nonlinear (logarithmic)
dependence of $E$ on parameters. For example, in the case $\Delta=\delta=0$
of greatest interest, the peak width (see Fig.~\ref{OutputSpectrum}f,
Fig.~\ref{EntanglementComparison}c) is not set by the small mechanical
linewidth $\Gamma$, unlike for the output spectrum discussed above.
Thus, the values of the entanglement rate $\Gamma_{E}$ for the parameters
explored here are much larger, of the order $\kappa\cdot{\rm max}_{\omega}E[\omega]$.
For increasing temperatures, $\Gamma_{{\rm E}}$ decreases, but only
slowly, indicating robust entanglement in this setup: $\Gamma_{E}\propto{\rm n}_{{\rm th}}^{-1}$,
with the prefactor set by the cooperativity $\mathcal{C}$ (Fig.~\ref{EntanglementComparison}d).

\section{Conclusions}

We have introduced an entanglement rate as a quantitative measure
for the CV entanglement production per unit time in setups involving
resonant modes. The definition is natural, in that it simply characterizes
the total entanglement between two beams within a time-interval of
size $T$, in the limit $T\rightarrow\infty$. In principle, it is
also more general than the Gaussian CV case studied in the present
manuscript.

Moreover, we have studied an optomechanical setup with one mechanical
and three optical modes that allows fully resonant production of optical
entanglement. The spectral density of entanglement and the overall
entanglement rate are optimized for different parameter choices. The
concept introduced here should be useful for analyzing setups in other
domains, like cavities with a Kerr medium \cite{KIMBLE,FUERST} or
microwave resonators with nonlinearities \cite{WALLRAFF,HUARD}.

\emph{Acknowledgement}. - We have benefited from discussions with
A.~Clerk, D.~Vitali and T.~Kiendl, as well as S.~Kessler, A.~Kronwald,
R.~Lauter, M.~Ludwig, V.~Peano, T. Paraiso, M.~Schmidt and T.~Weiss.
ZJD thanks FM for his kind hospitality in Erlangen. This work was
supported via an ERC Starting Grant OPTOMECH, via ITN cQOM, and via
the DARPA program ORCHID. ZJD acknowledges support from the National
Natural Science Foundation of China under Grants No. 11104353 and
No. 11574398.

\section*{Appendix}

\subsection{Wave Packet Basis: General Scheme}

\label{AppendixWavePacketBasisGeneral}In the main text, we claimed
that our general entanglement rate can be connected to an approach
of frequency-filtering and decomposing the filtered beams into wave
packets. The natural way to make these notions precise is by using
the Wannier-basis of wave packets $f_{m,n}(t)$ that live on a regular
grid both in time, $t_{n}=n\tau$, and in frequency-space, $\omega_{m}=m\delta\omega$,
where $\delta\omega=2\pi/\tau$. The Fourier-transform $F_{m,n}(\omega)\equiv\int e^{i\omega t}f_{m,n}(t)dt/\sqrt{2\pi}$
of these wave packets is nonzero only in the interval $\omega\in[(m-\frac{1}{2})\delta\omega,(m+\frac{1}{2})\delta\omega)$,
where $F_{m,n}(\omega)=\exp(i\omega t_{n})/\sqrt{\delta\omega}$.
The $F_{m,n}(\omega)$, and therefore also the $f_{m,n}(t)$, form
an orthonormal basis. That makes it possible to decompose uniquely
the output field into modes defined by these wave packets: $\hat{A}_{s}(t)=\sum_{m,n}\hat{a}_{m,n}^{(s)}f_{m,n}(t)$,
where $\hat{a}_{m,n}^{(s)}=\int f_{m,n}^{*}(t)\hat{A}_{s}(t)\,dt$
annihilates a photon in mode $(m,n)$, and $\left[\hat{a}_{m,n}^{(s)},\hat{a}_{\tilde{m},\tilde{n}}^{(s')\dagger}\right]=\delta_{m,\tilde{m}}\delta_{n,\tilde{n}}\delta_{s,s'}$.
In a typical application with a stationary-state source producing
beams via parametric down-conversion or four-wave mixing, energy conservation
dictates that pairs of frequencies $\omega_{1}$ and $\omega_{2}$
are entangled only when they obey $\omega_{{\rm total}}=\omega_{1}+\omega_{2}$
(in the limit $\tau\rightarrow\infty$). In the rotating frame adopted
in the main text, $\omega_{{\rm total}}$ would be zero.

We now calculate the logarithmic negativity $E_{N}^{\tau}$ between
two wave packets of the two output fields, at the same time slot $t_{n}$
and in suitable frequency slots $\omega$ and $-\omega$. For simplicity,
we will assume a situation with purely inter-beam entanglement (no
intra-beam squeezing), as discussed at the end of section \ref{sec:EntanglementRate}.
A similar construction would apply to the more general case, even
though then it would become necessary to consider the entanglement
between two wave packets at $\omega$ and $-\omega$ of beam 1 with
two wave packets at $\omega$ and $-\omega$ of beam 2. 

To prepare for our definition of the entanglement rate, we have to
discuss the dependence of $E_{N}^{\tau}$ on the filter time $\tau$,
especially for the limit $\tau\gg\tau_{c}$, where $\tau$  is much
longer than the physical correlation time $\tau_{c}$ of the source
($\tau_{c}^{-1}$ is the width of the spectral peaks). First, this
ensures that there will be no correlations between wave packets located
at different time slots, such that we capture the full amount of entanglement.
Second, we will now explain in this wave packet picture why $E_{N}^{\tau}$
tends to a well-defined limit for $\tau\rightarrow\infty$, which
is consistent with direct calculations\cite{GENES}.

Consider enlarging the filter time $\tau$ by a factor $M$, shrinking
the filter frequency interval by $\delta\omega'=\delta\omega/M$.
In effect, the new wave packets of size $\tau'=M\tau$ encompass $M$
of the old wave packets. That this coarse-graining keeps the correlators,
and thus the entanglement $E_{N}^{\tau}$, unchanged can be understood
already from a very simple consideration. Take a suitably normalized
sum of $M$ operators, $\hat{X}=M^{-1/2}\sum_{n=1}^{M}\hat{X}_{n}$
and likewise $\hat{Y}=M^{-1/2}\sum_{n=1}^{M}\hat{Y}_{n}$. Then the
correlator between these ``averaged'' operators will equal the original
correlator: $\left\langle \hat{X}\hat{Y}\right\rangle =\left\langle \hat{X}_{n}\hat{Y}_{n}\right\rangle $.
This holds provided there have been no cross-correlations and $\left\langle \hat{X}_{n}\hat{Y}_{n}\right\rangle $
is independent of $n$. The same logic applies to our case, where
the new modes are properly normalized averages over the old modes:
$\hat{a}{}_{m',n'}^{'}=\sum_{n}K_{m'}(n-n'M)\hat{a}_{m,n}$, with
$K$ encoding the overlap between the two basis sets. The detailed
structure of $K$ is not important for our argument, but essentially
$K$ is nonzero only in a range of size $M$, for $\left|n-n'M\right|\lesssim M$,
and it is normalized: $\sum_{n}\left|K_{m'}(n-n'M)\right|^{2}=1$.
As shown in the next section, this ensures a well-defined limit $E_{N}^{\tau\rightarrow\infty}$.

The entanglement rate for a given frequency slot $m$ may now be defined
naturally as $E_{N,m}^{\tau}/\tau$, the ratio between the entanglement
$E_{N,m}^{\tau}$ contained in a pair of wave packets at that frequency
and the time $\tau$ between those wave packets. (Here $m$ would
denote the index for beam 1 to which corresponds uniquely an index
$m_{2}$ for beam 2, as explained above.) As the logarithmic negativity
is additive, it makes sense to add up $E_{N,m}^{\tau}$ for all the
small frequency intervals into which the emission of the source has
been decomposed.

We have checked by direct calculation that $\lim_{\tau\rightarrow\infty}E_{N,m}^{\tau}$
defined here co-incides with the $E[\omega]$ defined in the main
text (where $\omega=m\delta\omega$ is kept fixed in the limit $\tau\rightarrow\infty$,
by adjusting $m$). This is despite the fact that $E_{N,m}^{\tau}$
is defined with respect to Wannier basis modes, while $E[\omega]$
was defined from the Fourier components of the field on a finite time-interval
of length $T\gg\tau_{c}$. 

We thus have arrived again (in a different route), at the total entanglement
rate:

\begin{equation}
\Gamma_{E}\equiv\lim_{\tau\rightarrow\infty}\sum_{m}\frac{E_{N,m}^{\tau}}{\tau}=\int_{-\infty}^{+\infty}\frac{d\omega}{2\pi}E[\omega]\,.
\end{equation}
We have used $\delta\omega=2\pi/\tau$ to convert the sum into an
integral. In the limit $\tau\rightarrow\infty$, the details of the
Wannier basis have become unimportant. $E[\omega]$ can now be calculated
using any filter function.

\subsection{More details on the Wannier basis}

In this section, we present some more technical details on our wave
packet based discussion of the entanglement rate. As defined in the
main text and mentioned in the preceding section, we start from a
complete orthonormal basis of packets that are localized both on a
time-grid ($t_{n}=n\tau$) and a frequency-grid ($\omega_{m}=m\delta\omega$),
where $\tau=2\pi/\delta\omega$. In frequency space, these basis functions
are defined as

\begin{equation}
F_{m,n}(\omega)=\frac{1}{\sqrt{\delta\omega}}e^{i\omega t_{n}}\,,
\end{equation}
if $(m-1/2)\delta\omega\leq\omega<(m+1/2)\delta\omega$, and $F_{m,n}(\omega)=0$
otherwise. In time-space, this corresponds to the well-known Wannier-basis
of sinc-shaped wave packets,

\[
f_{m,n}(t)=\int F_{m,n}(\omega)e^{-i\omega t}\frac{d\omega}{\sqrt{2\pi}}=f_{m}(t-t_{n})\,,
\]
where

\begin{equation}
f_{m}(t)=\frac{1}{\sqrt{\tau}}e^{-i\omega_{m}t}\frac{\sin(\delta\omega t/2)}{\delta\omega t/2}\,.
\end{equation}

Upon temporal coarse-graining, we combine $M$ old wave packets into
one new one, and at the same time the frequency-resolution becomes
refined: $\tau'=M\tau$ and $\delta\omega'=\delta\omega/M$. Thus,
the new frequency index $m'$ can be thought of as a combination of
the old index $m$ and another index $l=0\ldots N-1$ that splits
the old frequency interval into $M$ pieces. The interval belonging
to $m'$ is thus: $\omega\in[(m-1/2)\delta\omega+l\delta\omega',(m-1/2)\delta\omega+(l+1)\delta\omega'[$.
The definition of $F'_{m',n'}(\omega)$ on this interval reads like
the old one, except for the obvious replacements: $F'_{m',n'}(\omega)=e^{i\omega t'_{n'}}/\sqrt{\delta\omega'}$,
with $t'_{n'}\equiv n'\tau'=n'M\tau$. Both the old and the new basis
are complete.

We now want to obtain the overlap integrals that relate the old basis
to the new one. We find:

\begin{equation}
\left\langle m',n'|m,n\right\rangle \equiv\int F^{'*}{}_{m',n'}(\omega)F_{m,n}(\omega)d\omega=K_{m'}(n-n'M)\,,
\end{equation}
with

\begin{equation}
K_{m'}(k)=\begin{cases}
k=0: & \frac{1}{\sqrt{M}}\\
k\neq0: & \sqrt{M}\frac{e^{i\frac{2\pi}{M}k}-1}{2\pi ik}(-1)^{k}e^{i\frac{2\pi}{M}lk}
\end{cases}
\end{equation}
Note that the overlap $K$ obviously depends on the frequency index
$m'$ only via the refinement index $l$ in $m'=(m,l)$, and that
$K=0$ if we were to calculate the overlap for any $m'$ that is not
part of the original frequency interval defined by $m$. The shape
of the overlap as a function of the temporal distance $n-n'M$ is
that of a sinc function with a decay scale set by $M$. One can confirm
that the overlap matrix elements calculated here are normalized, $\sum_{n}\left|K_{m'}(n-n'M)\right|^{2}=1$.

When we think of the situation with two entangled beams, we imagine
each of them is defined by its own fluctuating output field, $\hat{A}_{\sigma}(t)$,
where $\sigma=1,2$ denotes the beam. Each of those can be decomposed
into the kind of Wannier basis defined here, and we choose the same
filter time $\tau$ for each of them. Regarding the frequencies, we
want to exploit the fact that in a typical steady-state situation
(like in parametric down-conversion or four-wave mixing), it is pairs
of frequencies that are entangled. Thus, to each frequency $\omega_{1}$
of beam 1 belongs an entangled frequency $\omega_{2}$ of beam 2 (with
$\omega_{{\rm total}}=\omega_{1}+\omega_{2}$). To simplify the subsequent
notation, we want to shift and revert the frequency scale of beam
2 such that the two mutually entangled frequencies are always both
denoted by the index $m$. In other words, while $\omega_{1}=m\delta\omega$,
we have $\omega_{2}=\omega_{{\rm total}}-m\delta\omega$. We note
that, after going into a rotating frame that makes the Hamiltonian
time-independent (as in the main text), we obtain $\omega_{{\rm total}}=0$.
In addition, it turns out that due to this matching between opposite
frequencies, the basis functions for beam 2 have to be changed accordingly,
and the basis transformation for beam 2 is effected by $K^{*}$ instead
of $K$.

We now want to confirm (as indicated in the main text), that such
a coarse-graining does not change the entanglement $E$ between wave
packets, provided we are already in the regime of sufficiently large
filter time, $\tau\gg\tau_{c}$. To this end, we just have to show
that the correlators between modes do not change upon coarse-graining,
i.e. we want to show in particular

\begin{equation}
\left\langle \hat{a}_{m,n}^{(1)}\hat{a}_{m,n}^{(2)}\right\rangle =\left\langle \hat{a}_{m',n'}^{'(1)}\hat{a}_{m',n'}^{'(2)}\right\rangle \label{eq:correlators_equal}
\end{equation}
where the modes on the right-hand side are the temporally coarse-grained
ones, and $\omega_{m'}$ lies within the interval defined by $m$.
The precise location of $\omega_{m'}$ within that interval will not
matter, since $\delta\omega\ll1/\tau_{c}$, so the spectrum of the
source is already flat on that scale. In addition, we note that, for
the steady-state situation we assume here, neither the left-hand-side
nor the right-hand-side of (\ref{eq:correlators_equal}) actually
depend on the time-point $n$ or $n'$, respectively.

Employing the overlap calculated above, we find:

\begin{eqnarray}
 & \left\langle \hat{a}_{m',n'}^{'(1)}\hat{a}_{m',n'}^{'(2)}\right\rangle =\nonumber \\
 & \sum_{n_{1},n_{2}}K_{m'}(n_{1}-n'M)K_{m'}^{\ast}(n_{2}-n'M)\left\langle \hat{a}_{m,n_{1}}^{(1)}\hat{a}_{m,n_{2}}^{(2)}\right\rangle =\nonumber \\
 & \sum_{n}\left|K_{m'}(n-n'M)\right|^{2}\left\langle \hat{a}_{m,n}^{(1)}\hat{a}_{m,n}^{(2)}\right\rangle =\left\langle \hat{a}_{m,n}^{(1)}\hat{a}_{m,n}^{(2)}\right\rangle 
\end{eqnarray}
Here we have used that $\hat{a}_{m',n'}^{'(1)}=\sum_{n_{1}}K_{m'}(n_{1}-n'M)\hat{a}_{m,n_{1}}^{(1)}$
and $\hat{a}_{m',n'}^{'(2)}=\sum_{n_{2}}K_{m'}^{\ast}(n_{2}-n'M)\hat{a}_{m,n_{2}}^{(2)}$.
In going to the second line, we exploited the fact that different
time slots are already uncorrelated ($\tau\gg\tau_{c}$). We also
used the normalization of $K$ in the last step. 

Regarding other correlators, such as $\left\langle \hat{a}_{m,n}^{(1)}\hat{a}_{m,n}^{(2)\dagger}\right\rangle $,
we can say the following: For the model discussed in the main text,
they can be confirmed to be zero by explicit calculation. For more
general models, where however the Hamiltonian can still be brought
to a time-independent form by a proper choice of rotating frame, we
find that stationarity dictates that $\left\langle \hat{a}^{(1)}(\omega)\left(\hat{a}^{(2)\dagger}\right)(\omega')\right\rangle \propto\delta(\omega+\omega')$,
and likewise for all other correlators. In evaluating a correlator
of amplitudes in the Wannier-basis, like $\left\langle \hat{a}_{m,n}^{(1)}\hat{a}_{m,n}^{(2)\dagger}\right\rangle $,
we are effectively looking at correlators of the type $\left\langle \hat{a}^{(1)}(\omega)\left(\hat{a}^{(2)}(\omega')\right)^{\dagger}\right\rangle $,
with $\omega'\approx-\omega$ {[}by our definition given above, $m$
enters the frequency $\omega_{2}$ with an opposite sign{]}. Since
$\left(\hat{a}^{(2)}(\omega')\right)^{\dagger}=\left(\hat{a}^{(2)\dagger}\right)(-\omega')$,
that correlator equals $\left\langle \hat{a}^{(1)}(\omega)\left(\hat{a}^{(2)\dagger}\right)(-\omega')\right\rangle \propto\delta(\omega-\omega')$.
By virtue of $\omega'\approx-\omega$, this is zero automatically.
(Note that formally we have to exclude the single point $\omega=0$
in this argument, which would physically equal the incoming laser
frequency in the case of four-wave mixing, and the corresponding small
frequency slot.) For the correlators involving quantities of the same
beam, it is rather correlators of the type $\left\langle \hat{a}_{m,n}^{(1)}\hat{a}_{m,n}^{(1)\dagger}\right\rangle $
that are nonzero, while correlators like $\left\langle \hat{a}_{m,n}^{(1)}\hat{a}_{m,n}^{(1)}\right\rangle $
are zero, and the proof is in analogy to what was shown above. In
summary, the entanglement $E$ will go to a well-defined limit as
$\tau\rightarrow\infty$ ($\tau\gg\tau_{c}$). (As mentioned above,
in taking this limit, we assume the frequency index $m$ is re-adjusted
such that the center frequency of the corresponding slot is held fixed
at some given $\omega$, thus arriving at $E[\omega]$ in the limit
$\tau\rightarrow\infty$.) The actual calculation of $E[\omega]$
is discussed below.

\subsection{Hamiltonian and implementation with three coupled optomechanical
cells}

In this section, we give a derivation of the Hamiltonian (2) and discuss
its implementation. We consider three equidistant optical modes with
a splitting $J$ coupled to the same mechanical mode $\hat{b}$ with
frequency $\Omega$ via radiation pressure. One of the optical modes
(here called $a_{0}$) is driven by an external laser at frequency
$\omega_{L}$. Such a setup in general can be described by the following
Hamiltonian ($\hbar=1$), where for brevity we do not display the
coupling to the dissipative environment:

\begin{equation}
\hat{H}=\sum_{q=\pm,0}\omega_{q}\hat{a}_{q}^{\dagger}\hat{a}_{q}+\Omega\hat{b}^{\dagger}\hat{b}-\sum_{q,q'=\pm,0}g_{q',q}^{(0)}\hat{a}_{q'}^{\dagger}\hat{a}_{q}(\hat{b}+\hat{b}^{\dagger})+\Lambda(\hat{a}_{0}e^{i(\omega_{L}t+\phi)}+{\rm h.c.})\label{fullHam}
\end{equation}
Here $\omega_{\pm}=\omega_{0}\pm J$, and $g_{q',q}^{(0)}$ represents
the generic (Hermitian) optomechanical coupling matrix. We now assume
that the frequency mismatch $\delta\equiv\Omega-J$ is much smaller
than the mechanical frequency, i.e., $\left|\delta\right|\ll\Omega$
and that the mechanical sidebands are resolved, i.e., $\kappa\ll\Omega$,
where $\kappa$ is the optical damping rate. After transforming to
the rotating frame with respect to $\hat{H}_{0}=(\omega_{L}+J)\hat{a}_{+}^{\dagger}\hat{a}_{+}+(\omega_{L}-J)\hat{a}_{-}^{\dagger}\hat{a}_{-}+\omega_{L}\hat{a}_{0}^{\dagger}\hat{a}_{0}+J\hat{b}^{\dagger}\hat{b}$,
and neglecting rapidly oscillating terms rotating at $\pm J$ by a
rotating wave approximation (RWA), we find

\begin{equation}
\hat{H}_{{\rm {\rm RWA}}}=\sum_{q=\pm,0}-\Delta\hat{a}_{q}^{\dagger}\hat{a}_{q}+\delta\hat{b}^{\dagger}\hat{b}-\{(g_{+,0}^{(0)}\hat{a}_{+}^{\dagger}\hat{a}_{0}+g_{0,-}^{(0)}\hat{a}_{0}^{\dagger}\hat{a}_{-})\hat{b}+h.c\}+\Lambda(\hat{a}_{0}e^{i\phi}+\hat{a}_{0}^{\dagger}e^{-i\phi})
\end{equation}
where $\Delta=\omega_{L}-\omega_{0}$ is the laser drive detuning.
For a sufficiently strong laser drive, we can linearize the dynamics
by replacing $\hat{a}_{0}$ by a complex number $\alpha=\frac{-i\Lambda e^{-i\phi}}{-i\triangle+\kappa/2}$.
Thus we find the Hamiltonian (2) as given in the main text

\begin{equation}
\hat{H}_{{\rm lin}}=-\triangle(\hat{a}_{+}^{\dagger}\hat{a}_{+}+\hat{a}_{-}^{\dagger}\hat{a}_{-})+\delta\hat{b}^{\dagger}\hat{b}-\frac{g}{2}\{(\hat{a}_{+}^{\dagger}+\hat{a}_{-})\hat{b}+h.c\},
\end{equation}
provided that the couplings to the ``$+$'' and ``$-$'' mode
turn out to be equal, i.e. $g/2\equiv g_{+,0}^{(0)}\alpha=g_{0,-}^{(0)}\alpha$.
Without loss of generality, we assume that $g$ is real-valued. The
RWA made above is valid when $\left|g\right|\ll\Omega$. It may be
unavoidable that there is a small asymmetry in the two optomechanical
coupling strengths, but since the entanglement generation involves
both transitions simultaneously this does not make a crucial difference
(we have confirmed this for the output entanglement which we discuss).
A more detailed account on some further aspects of asymmetric optomechanical
coupling strengths in the context of this kind of Hamiltonian may
be found in \cite{WANG,TIAN}, where it is pointed out that for the
case of intra-cavity entanglement (which is not our concern here)
having such an asymmetry can actually be beneficial.

We now turn to deriving the Hamiltonian (\ref{fullHam}) for a concrete
setup consisting of three coupled optomechanical cells. In each of
the cells, we assume a standard (local) coupling between photons and
phonons. The microscopic Hamiltonian reads ($\hbar=1$)

\begin{equation}
\hat{H}_{{\rm cell}}=\sum_{l=1}^{3}\{\omega_{0}\hat{a}_{l}^{\dagger}\hat{a}_{l}+\Omega\hat{b}_{l}^{\dagger}\hat{b}_{l}-g_{0}\hat{a}_{l}^{\dagger}\hat{a}_{l}(\hat{b}_{l}^{\dagger}+\hat{b}_{l})\}-(K_{1}\hat{a}_{1}^{\dagger}\hat{a}_{2}+K_{2}\hat{a}_{2}^{\dagger}\hat{a}_{3}+h.c.)
\end{equation}
where the index $l=1,2,3$ runs over the three sites. The second term
describes photon tunneling between different sites. It can be taken
into account by introducing optical normal modes, as defined by $\hat{a}_{\pm}=(\frac{K_{1}\hat{a}_{1}+K_{2}\hat{a}_{3}}{\sqrt{K_{1}^{2}+K_{2}^{2}}}\mp\hat{a}_{2})/\sqrt{2}$,
$\hat{a}_{0}=\frac{K_{2}\hat{a}_{1}-K_{1}\hat{a}_{3}}{\sqrt{K_{1}^{2}+K_{2}^{2}}}$
with eigenfrequencies $\omega_{\pm}=\omega_{0}\pm\sqrt{K_{1}^{2}+K_{2}^{2}}$,
$\omega_{0}$ respectively. In terms of the normal modes, the Hamiltonian
can be written as

\begin{equation}
\begin{array}{c}
\hat{H}_{{\rm cell}}=\sum_{q=\pm,0}\omega_{q}\hat{a}_{q}^{\dagger}\hat{a}_{q}+\sum_{l=1}^{3}\Omega\hat{b}_{l}^{\dagger}\hat{b}_{l}-\frac{g_{0}}{2}(\hat{a}_{-}-\hat{a}_{+})^{\dagger}(\hat{a}_{-}-\hat{a}_{+})(\hat{b}_{2}^{\dagger}+\hat{b}_{2})\\
-\frac{g_{0}}{2(K_{1}^{2}+K_{2}^{2})}(K_{1}\hat{a}_{+}+K_{1}\hat{a}_{-}+\sqrt{2}K_{2}\hat{a}_{0})^{\dagger}(K_{1}\hat{a}_{+}+K_{1}\hat{a}_{-}+\sqrt{2}K_{2}\hat{a}_{0})(\hat{b}_{1}^{\dagger}+\hat{b}_{1})\\
-\frac{g_{0}}{2(K_{1}^{2}+K_{2}^{2})}(K_{2}\hat{a}_{+}+K_{2}\hat{a}_{-}-\sqrt{2}K_{1}\hat{a}_{0})^{\dagger}(K_{2}\hat{a}_{+}+K_{2}\hat{a}_{-}-\sqrt{2}K_{1}\hat{a}_{0})(\hat{b}_{3}^{\dagger}+\hat{b}_{3})
\end{array}
\end{equation}
This Hamiltonian takes essentially the same form as the Hamiltonian
(\ref{fullHam}) given above. We assume that $\delta=\Omega-\sqrt{K_{1}^{2}+K_{2}^{2}}\ll\Omega$
and $\kappa\ll\Omega$, transform to a rotating frame with $\hat{H_{0}}=\sum_{q=\pm,0}(\omega_{q}-\omega_{0})\hat{a}_{q}^{\dagger}\hat{a}_{q}+\sum_{l=1}^{3}\sqrt{K_{1}^{2}+K_{2}^{2}}\hat{b}_{l}^{\dagger}\hat{b}_{l}$,
and apply a RWA to find

\begin{equation}
\hat{H}_{{\rm eff}}=\sum_{q=\pm,0}\omega_{0}\hat{a}_{q}^{\dagger}\hat{a}_{q}+\sum_{l=1}^{3}\delta\hat{b}_{l}^{\dagger}\hat{b}_{l}-g_{0}\frac{K_{1}K_{2}}{K_{1}^{2}+K_{2}^{2}}\{(\hat{a}_{+}^{\dagger}\hat{a}_{0}+\hat{a}_{-}\hat{a}_{0}^{\dagger})\frac{\hat{b}_{1}-\hat{b}_{3}}{\sqrt{2}}+h.c.\}
\end{equation}
After adding an external laser drive for the $\hat{a}_{0}$ mode,
moving into a frame rotating with the drive frequency and applying
standard linearization, this Hamiltonian is identical in form to the
Hamiltonian (2) given in the main text. The relevant mechanical normal
mode is given by $\hat{b}=(\hat{b}_{1}-\hat{b}_{3})/\sqrt{2}.$ Note
that the design is quite flexible in that it also applies to setups
with unbalanced hopping rates $K_{1}\neq K_{2}$ and that, in principle,
a mechanical mode coupled only to either the first or the third site
would suffice. The $\hat{a}_{0}$ mode may be driven through an additional
channel, provided that the decay rate into this is sufficiently small
so as to ensure that the entangled photon pairs dominantly decay into
another, outcoupling waveguide.

As pointed out in the main text, a similar configuration of levels
may be realized in optomechanical membrane-in-the-middle setups. In
such setups, the frequencies of the transverse normal optical modes
of the cavity depend on the membrane position and tilt angle \cite{HARRIS3}.
These parameters may be tuned so as to create triplet equidistant
optical modes. The frequency separation can be comparable to the frequency
of a vibrational mode of the membrane, which could be matched to the
optical splitting by applying the optical spring effect \cite{MARQUARDTREVIEW}.
In such a configuration, the (linear) optomechanical coupling strengths
are set by the slopes of the optical bands \cite{HARRIS2}. As pointed
out above, even if these turn out to be different, that will not impact
our scheme in any important way.

\subsection{Effective optical model}

Next, we derive the effective optical model, i.e. the model obtained
after integrating out the mechanics. We will discuss how it captures
some essential features of the full model. Assuming a large frequency
mismatch $\delta\gg\kappa,\,\Gamma,\,g,\,|\Delta|$, and low temperatures,
i.e., $\sqrt{n_{{\rm th}}+1}<g/\kappa$, we can adiabatically eliminate
the mechanical mode and the mechanical bath. This can be accomplished
by a polaron transformation $\hat{H}_{{\rm opt}}=e^{\hat{S}}\hat{H}_{{\rm lin}}e^{-\hat{S}}$
with $\hat{S}=\frac{g}{2(\delta+\Delta)}(\hat{a}_{+}^{\dagger}\hat{b}-\hat{a}_{+}\hat{b}^{\dagger})+\frac{g}{2(\delta-\Delta)}(\hat{a}_{-}\hat{b}-\hat{a}_{-}^{\dagger}\hat{b}^{\dagger})$.
Thus, we find

\begin{equation}
\begin{array}{c}
\hat{H}_{{\rm opt}}=-\triangle(\hat{a}_{+}^{\dagger}\hat{a}_{+}+\hat{a}_{-}^{\dagger}\hat{a}_{-})+\delta\hat{b}^{\dagger}\hat{b}-(\frac{g^{2}}{8(\delta+\triangle)}+\frac{g^{2}}{8(\delta-\triangle)})(\hat{a}_{+}^{\dagger}\hat{a}_{-}^{\dagger}+\hat{a}_{+}\hat{a}_{-})\\
-\frac{g^{2}}{4(\delta+\triangle)}\hat{a}_{+}^{\dagger}\hat{a}_{+}-\frac{g^{2}}{4(\delta-\triangle)}\hat{a}_{-}^{\dagger}\hat{a}_{-}-(\frac{g^{2}}{4(\delta+\triangle)}-\frac{g^{2}}{4(\delta-\triangle)})\hat{b}^{\dagger}\hat{b}\\
\approx-\triangle(\hat{a}_{+}^{\dagger}\hat{a}_{+}+\hat{a}_{-}^{\dagger}\hat{a}_{-})+\delta\hat{b}^{\dagger}\hat{b}-\frac{g^{2}}{4\delta}(\hat{a}_{+}^{\dagger}\hat{a}_{+}+\hat{a}_{-}^{\dagger}\hat{a}_{-}+\hat{a}_{+}^{\dagger}\hat{a}_{-}^{\dagger}+\hat{a}_{+}\hat{a}_{-}).
\end{array}
\end{equation}
where, in the last step, we explicitly used that $|\Delta|\ll\delta$.
The optical modes are now decoupled from the mechanical mode and we
obtain a closed set of quantum Langevin equations for the optical
modes

\begin{equation}
\begin{array}{c}
\dot{\hat{a}}_{+}=i(\triangle+\frac{g^{2}}{4\delta})\hat{a}_{+}+i\frac{g^{2}}{4\delta}\hat{a}_{-}^{\dagger}-\frac{\kappa}{2}\hat{a}_{+}-\sqrt{\kappa}\hat{a}_{+,in}(t)\\
\dot{\hat{a}}_{+}^{\dagger}=-i(\triangle+\frac{g^{2}}{4\delta})\hat{a}_{+}^{\dagger}-i\frac{g^{2}}{4\delta}\hat{a}_{-}-\frac{\kappa}{2}\hat{a}_{+}^{\dagger}-\sqrt{\kappa}\hat{a}_{+,in}^{\dagger}(t)\\
\dot{\hat{a}}_{-}=i(\triangle+\frac{g^{2}}{4\delta})\hat{a}_{-}+i\frac{g^{2}}{4\delta}\hat{a}_{+}^{\dagger}-\frac{\kappa}{2}\hat{a}_{-}-\sqrt{\kappa}\hat{a}_{-,in}(t)\\
\dot{\hat{a}}_{-}^{\dagger}=-i(\triangle+\frac{g^{2}}{4\delta})\hat{a}_{-}^{\dagger}-i\frac{g^{2}}{4\delta}\hat{a}_{+}-\frac{\kappa}{2}\hat{a}_{-}^{\dagger}-\sqrt{\kappa}\hat{a}_{-,in}^{\dagger}(t)
\end{array}
\end{equation}
where $\left\langle \hat{a}_{q,in}\right\rangle =0$, $\left\langle \hat{a}_{q,in}^{\dagger}(t)\hat{a}_{q',in}(t')\right\rangle =0$,
$\left\langle \hat{a}_{q,in}(t)\hat{a}_{q',in}^{\dagger}(t')\right\rangle =\delta_{qq'}\delta(t-t')$
with $q,q'=\pm.$ The output fields are related by $\hat{a}_{\pm,out}(t)=\sqrt{\kappa}\hat{a}_{\pm}(t)+\hat{a}_{\pm,in}(t)$.
For notational convenience, we introduce the operator vectors $A=(\begin{array}{cccc}
\hat{a}_{+} & \hat{a}_{+}^{\dagger} & \hat{a}_{-} & \hat{a}_{-}^{\dagger}\end{array})$$^{T}$ and $A_{in}=(\begin{array}{cccc}
\hat{a}_{+,in} & \hat{a}_{+,in}^{\dagger} & \hat{a}_{-,in} & \hat{a}_{-,in}^{\dagger}\end{array})$$^{T}$. The quantum Langevin equations can be cast in the following
compact form

\begin{equation}
\dot{A}=MA-\sqrt{\kappa}A_{in}
\end{equation}
where

\[
M=\left(\begin{array}{cccc}
i(\triangle+\frac{g^{2}}{4\delta})-\frac{\kappa}{2} & 0 & 0 & i\frac{g^{2}}{4\delta}\\
0 & -i(\triangle+\frac{g^{2}}{4\delta})-\frac{\kappa}{2} & -i\frac{g^{2}}{4\delta} & 0\\
0 & i\frac{g^{2}}{4\delta} & i(\triangle+\frac{g^{2}}{4\delta})-\frac{\kappa}{2} & 0\\
-i\frac{g^{2}}{4\delta} & 0 & 0 & -i(\triangle+\frac{g^{2}}{4\delta})-\frac{\kappa}{2}
\end{array}\right).
\]
The system is stable only if all eigenvalues of $M$ have non-positive
real parts. The boundary of stability is located where one of the
real parts becomes zero, i.e., $-\frac{\kappa}{2}+\sqrt{-(\frac{g^{2}}{2\delta}+\triangle)\triangle}=0$,
or equivalently when $(\Delta+g^{2}/(2\delta))\Delta+\kappa^{2}/4=0$.
This corresponds to the dashed line in Fig. \ref{StabilityDiagram}a.
Note that the lower quadrants of the stability diagram are essentially
identical due to inversion symmetry with respect to the point $\delta=\Delta=0$.

We can solve the quantum Langevin equations in Fourier space. We define
the Fourier-transformed operators by $\hat{a}_{+}(\omega)=\intop_{-\infty}^{+\infty}\hat{a}_{+}(t)e^{i\omega t}dt$,
$\hat{a}_{+}^{\dagger}(-\omega)\equiv\intop_{-\infty}^{+\infty}\hat{a}_{+}^{\dagger}(t)e^{i\omega t}dt=(\hat{a}_{{\rm +}}^{\dagger})(\omega)$.
In the frequency domain, the input noise correlation reads $\left\langle \hat{a}_{q,in}(\omega)\hat{a}_{q',in}^{\dagger}(-\omega')\right\rangle =2\pi\delta_{qq'}\delta(\omega+\omega')$.
For convenience, we define the vectors $A_{in/out}(\omega)=(\begin{array}{cccc}
\hat{a}_{+,in/out}(\omega) & \hat{a}_{+,in/out}^{\dagger}(-\omega) & \hat{a}_{-,in/out}(\omega) & \hat{a}_{-,in/out}^{\dagger}(-\omega)\end{array})$$^{T}$. The solution of the Langevin equation can be written as $A_{out}(\omega)=S(\omega)A_{in}(\omega)$
with the scattering matrix

\begin{equation}
S(\omega)=\frac{\kappa}{M_{14}^{2}+(M_{11}+i\omega)(M_{22}+i\omega)}\left(\begin{array}{cccc}
M_{22}+i\omega & 0 & 0 & -M_{14}\\
0 & M_{11}+i\omega & M_{14} & 0\\
0 & -M_{14} & M_{22}+i\omega & 0\\
M_{14} & 0 & 0 & M_{11}+i\omega
\end{array}\right)+I
\end{equation}
where $I$ is the $4\times4$ identity matrix. The input noise correlations
$B_{in}(\omega,\omega')=\left\langle A_{in}(\omega)\otimes A_{in}^{{\rm T}}(\omega')\right\rangle $
are scattered into $B_{out}(\omega,\omega')=\left\langle A_{out}(\omega)\otimes A_{out}^{{\rm T}}(\omega')\right\rangle =S(\omega)B_{in}(\omega,\omega')S^{T}(\omega')$.

In the effective optical model, photons are only created in pairs.
The pair creation rate $\Gamma_{{\rm pairs}}$ is equal to the intensity
(photons/sec) in any of the two output streams. Due to the choice
of normalization in the input-output formalism, this is given by:

\begin{equation}
\begin{array}{c}
\left\langle \hat{a}_{\pm,out}^{\dagger}\hat{a}_{\pm,out}\right\rangle =(\frac{1}{2\pi})^{2}\int_{-\infty}^{+\infty}d\omega\int_{-\infty}^{+\infty}d\omega'\left\langle \hat{a}_{\pm,out}^{\dagger}(-\omega)\hat{a}_{\pm,out}(\omega')\right\rangle \\
=\frac{(g^{2}/4\delta)^{2}\kappa/2}{(\kappa/2)^{2}+(\triangle+\frac{g^{2}}{2\delta})\triangle}
\end{array}
\end{equation}
This result shows that, depending on the sign of $(\triangle+\frac{g^{2}}{2\delta})\triangle$,
we get an enhanced or decreased photon pair creation rate compared
with $\triangle=0$. In addition, since the denominator vanishes at
the boundary of stability, the photon pair creation rate diverges
there.

\subsection{Calculation of $E[\omega]$}

\label{AppendixCalcEntanglement}Here, we review the definition of
the logarithmic negativity and apply it to quantify the entanglement
of the filtered optical output fields, first for the effective optical
model introduced above, and then for the full model. By applying narrow
frequency filters, we select only single-frequency components of each
of the optical output fields. By energy conservation, correlations
only occur between the $\omega$-component of one of the fields (say
$\hat{a}_{+,out}$) and the $-\omega$-component of the other ($\hat{a}_{-,out}$).
The field operators for these two single-frequency output fields are
obtained as $\hat{a}_{+,\omega}(t)=\intop_{-\infty}^{t}f(t-s)\hat{a}_{+,out}(s)ds$,
$\hat{a}_{-,-\omega}(t)=\intop_{-\infty}^{t}g(t-s)\hat{a}_{-,out}(s)ds$
. For convenience, we now will choose Lorentzian filter functions
$f(t)=\sqrt{\frac{2}{\tau}}\theta(t)e^{-(\frac{1}{\tau}-i\omega)t}$,
$g(t)=\sqrt{\frac{2}{\tau}}\theta(t)e^{-(\frac{1}{\tau}+i\omega)t}$
with $\theta(t)$ the Heaviside step function, as opposed to the Wannier
basis used in the main text. In the limit of small bandwidth $1/\tau\rightarrow0$,
which is the only one we discuss, however, both the basis function
probability densities reduce to Dirac $\delta$-functions, and the
results will co-incide.

We can use the logarithmic negativity \cite{VIDAL} to characterize
the entanglement for the output light beams \cite{GENES}. In order
to evaluate it, we define the vector $u=(\begin{array}{cccc}
\hat{x}_{+,\omega}(t) & \hat{p}_{+,\omega}(t) & \hat{x}_{-,-\omega}(t) & \hat{p}_{-,-\omega}(t)\end{array})$$^{T}$ with $\hat{x}_{j}(t)=\frac{1}{\sqrt{2}}(\hat{a}_{j}(t)+\hat{a}_{j}^{\dagger}(t))$,
$\hat{p}_{j}(t)=\frac{1}{\sqrt{2}i}(\hat{a}_{j}(t)-\hat{a}_{j}^{\dagger}(t))$
($j=+,\omega$ or $j=-,-\omega$). The entanglement is determined
by the covariance matrix $V$ with matrix elements $V_{ij}=\frac{1}{2}\left\langle u_{i}u_{j}+u_{j}u_{i}\right\rangle $,
where the operators involved in this product are all taken at equal
times. Inserting the stationary solution, we find

\begin{equation}
V=\left(\begin{array}{cccc}
n_{+} & 0 & {\rm Re}(\xi) & {\rm Im}(\xi)\\
0 & n_{+} & {\rm Im}(\xi) & -{\rm Re}(\xi)\\
{\rm Re}(\xi) & {\rm Im}(\xi) & n_{-} & 0\\
{\rm Im}(\xi) & -{\rm Re}(\xi) & 0 & n_{-}
\end{array}\right),
\end{equation}
where 

\[
n_{+}=\int_{-\infty}^{+\infty}e^{-i\omega t}\left\langle \hat{a}_{+,{\rm out}}^{\dagger}(t)\hat{a}_{+,{\rm out}}(0)\right\rangle \,dt+1/2=|S_{14}(\omega)|^{2}+1/2\,,
\]

\[
n_{-}=\int_{-\infty}^{+\infty}e^{i\omega t}\left\langle \hat{a}_{-,{\rm out}}^{\dagger}(t)\hat{a}_{-,{\rm out}}(0)\right\rangle \,dt+1/2=|S_{14}(\omega)|^{2}+1/2\,,
\]
and 
\[
\xi=\int_{-\infty}^{+\infty}e^{i\omega t}\left\langle \hat{a}_{+,{\rm out}}(t)\hat{a}_{-,{\rm out}}(0)\right\rangle \,dt=S_{11}(\omega)S_{14}(-\omega)\,.
\]
The logarithmic negativity is defined as $E=max[0,-ln(2\eta_{-})]$
with $\eta_{-}=\frac{n_{+}+n_{-}}{2}-\sqrt{(\frac{n_{+}-n_{-}}{2})^{2}+|\xi|^{2}}$
being the smaller symplectic eigenvalue of the partial transpose of
matrix $V$. Choosing $\omega=0$, we plot $E[\omega=0]$ as a function
of $\triangle/\kappa$ in Fig.~\ref{StabilityDiagram}b.

To obtain the spectral density of entanglement in the\emph{ }full
model, we solve the following system of quantum Langevin equations,
which derive directly from the model Hamiltonian (main text), and
where dissipation and fluctuations have been taken into account using
the usual input-output formalism:

\begin{equation}
\begin{array}{c}
\dot{\hat{a}}_{+}=i\triangle\hat{a}_{+}+i\frac{g}{2}\hat{b}-\frac{\kappa}{2}\hat{a}_{+}-\sqrt{\kappa}\hat{a}_{+,in}(t)\\
\dot{\hat{a}}_{+}^{\dagger}=-i\triangle\hat{a}_{+}^{\dagger}-i\frac{g}{2}\hat{b}^{\dagger}-\frac{\kappa}{2}\hat{a}_{+}^{\dagger}-\sqrt{\kappa}\hat{a}_{+,in}^{\dagger}(t)\\
\dot{\hat{a}}_{-}=i\triangle\hat{a}_{-}+i\frac{g}{2}\hat{b}^{\dagger}-\frac{\kappa}{2}\hat{a}_{-}-\sqrt{\kappa}\hat{a}_{-,in}(t)\\
\dot{\hat{a}}_{-}^{\dagger}=-i\triangle\hat{a}_{-}^{\dagger}-i\frac{g}{2}\hat{b}-\frac{\kappa}{2}\hat{a}_{-}^{\dagger}-\sqrt{\kappa}\hat{a}_{-,in}^{\dagger}(t)\\
\dot{\hat{b}}=-i\delta\hat{b}+i\frac{g}{2}(\hat{a}_{+}+\hat{a}_{-}^{\dagger})-\frac{\Gamma}{2}\hat{b}-\sqrt{\Gamma}\hat{b}_{in}(t)\\
\dot{\hat{b}}^{\dagger}=i\delta\hat{b}^{\dagger}-i\frac{g}{2}(\hat{a}_{-}+\hat{a}_{+}^{\dagger})-\frac{\Gamma}{2}\hat{b}^{\dagger}-\sqrt{\Gamma}\hat{b}_{in}^{\dagger}(t)
\end{array}
\end{equation}
where, in addition, we have $\left\langle \hat{b}_{in}\right\rangle =0$,
$\left\langle \hat{b}_{in}^{\dagger}(t)\hat{b}_{in}(t')\right\rangle =n_{{\rm th}}\delta(t-t')$,
$\left\langle \hat{b}_{in}(t)\hat{b}_{in}^{\dagger}(t')\right\rangle =(n_{{\rm th}}+1)\delta(t-t')$,
with $n_{{\rm th}}=(\exp(\hbar\Omega/(k_{{\rm B}}T))-1)^{-1}$. For
the logarithmic negativity, we need to evaluate the same optical correlations
as above. The results are analytical, but too complicated to be reported
here. Simpler analytic results can be found for $\delta=\Delta=0$.
In this case we find $\eta_{-}(\omega,\delta,\triangle=0)=4\mathcal{C}(\mathcal{C}+n_{{\rm th}}+\frac{1}{2})+\frac{1}{2}-2\mathcal{C}\sqrt{1+4(\mathcal{C}+n_{{\rm th}}+\frac{1}{2})^{2}}$
with the cooperativity $\mathcal{C}=\frac{g^{2}}{\Gamma\kappa}$.
The entanglement $E[\omega,\delta,\triangle=0]$ is only determined
by $\mathcal{C}$ and $n_{{\rm th}}$ as given in the main text.

\subsection{Comparison to two- and three-mode schemes}

Finally, we compare the efficiency of our four-mode setup (3 optical,
1 vibrational) to two- and three-mode schemes that have been discussed
previously. We show that the four-mode setup is more efficient than
either of these schemes.

For a two-mode setup (1 optical, 1 vibrational) and with a resonant
laser drive (detuning $\Delta=0$), the Hamiltonian is given by $\hat{H}=\Omega\hat{b}^{\dagger}b-g(\hat{a}+\hat{a}^{\dagger})(\hat{b}+\hat{b}^{\dagger})$,
where $\hat{a}$ is an optical and $\hat{b}$ is a mechanical mode
(see \cite{MARQUARDTREVIEW} and \cite{GENES} for the case of drive
at the red mechanical sideband). The filtered correlations that quantify
the entanglement between the Stokes and anti-Stokes mechanical sidebands
are then given by

\begin{equation}
\begin{array}{c}
n_{+}=\frac{\kappa^{2}g^{4}}{(\Omega^{2}+(\kappa/2)^{2})^{2}(\Gamma/2)^{2}}+\frac{2n_{{\rm th}}\kappa g^{2}}{(\Omega^{2}+(\kappa/2)^{2})(\Gamma/2)}+\frac{1}{2}\\
n_{-}=\frac{\kappa^{2}g^{4}}{(\Omega^{2}+(\kappa/2)^{2})^{2}(\Gamma/2)^{2}}+\frac{2(n_{{\rm th}}+1)\kappa g^{2}}{(\Omega^{2}+(\kappa/2)^{2})(\Gamma/2)}+\frac{1}{2}\\
\xi=-\frac{\kappa^{2}g^{4}}{(\Omega^{2}+(\kappa/2)^{2})^{2}(\Gamma/2)^{2}}-\frac{(2n_{{\rm th}}+1)\kappa g^{2}}{(\Omega^{2}+(\kappa/2)^{2})(\Gamma/2)}
\end{array}
\end{equation}
where we have assumed that $\Omega\gg\Gamma$. For our model, with
resonant drive $\Delta=0$, we find

\begin{equation}
\begin{array}{c}
n_{+}=\frac{\kappa\Gamma n_{{\rm th}}(g/2)^{2}}{[(-\omega+\delta)^{2}+(\Gamma/2)^{2}](\omega^{2}+(\kappa/2)^{2})}+\frac{\kappa^{2}(g/2)^{4}}{[(-\omega+\delta)^{2}+(\Gamma/2)^{2}](\omega^{2}+(\kappa/2)^{2})^{2}}+\frac{1}{2}\\
n_{-}=\frac{\kappa\Gamma(n_{{\rm th}}+1)(g/2)^{2}}{[(-\omega+\delta)^{2}+(\Gamma/2)^{2}](\omega^{2}+(\kappa/2)^{2})}+\frac{\kappa^{2}(g/2)^{4}}{[(-\omega+\delta)^{2}+(\Gamma/2)^{2}](\omega^{2}+(\kappa/2)^{2})^{2}}+\frac{1}{2}\\
\xi=-\frac{\kappa^{2}(g/2)^{4}}{[(-\omega+\delta)^{2}+(\Gamma/2)^{2}](\omega^{2}+(\kappa/2)^{2})^{2}}+\frac{-\kappa\Gamma(n_{{\rm th}}+1/2)(g/2)^{2}+i(-\omega+\delta)\kappa(g/2)^{2}}{[(-\omega+\delta)^{2}+(\Gamma/2)^{2}](\omega^{2}+(\kappa/2)^{2})}
\end{array}
\end{equation}
In order to evaluate the entanglement between the Stokes and anti-Stokes
sidebands, we set $\omega=\delta$. This yields

\begin{equation}
\begin{array}{c}
n_{+}=\frac{2\kappa n_{{\rm th}}(g/2)^{2}}{(\Gamma/2)(\delta^{2}+(\kappa/2)^{2})}+\frac{\kappa^{2}(g/2)^{4}}{(\Gamma/2)^{2}(\delta^{2}+(\kappa/2)^{2})^{2}}+\frac{1}{2}\\
n_{-}=\frac{2\kappa(n_{{\rm th}}+1)(g/2)^{2}}{(\Gamma/2)(\delta^{2}+(\kappa/2)^{2})}+\frac{\kappa^{2}(g/2)^{4}}{(\Gamma/2)^{2}(\delta^{2}+(\kappa/2)^{2})^{2}}+\frac{1}{2}\\
\xi=-\frac{\kappa^{2}(g/2)^{4}}{(\Gamma/2)^{2}(\delta^{2}+(\kappa/2)^{2})^{2}}-\frac{\kappa(2n_{{\rm th}}+1)(g/2)^{2}}{(\Gamma/2)(\delta^{2}+(\kappa/2)^{2})}
\end{array}
\end{equation}
By comparing the coherent parts of $n_{+}$ and $n_{-}$ (the terms
that do not depend on $n_{{\rm th}}$ and are useful for optical entanglement),
we find that, in our setup, the photon pair creation rate is enhanced
by a factor $\sim\frac{\Omega^{4}}{(\delta^{2}+(\kappa/2)^{2})^{2}}$.

The main benefit of our setup in comparison to three-mode schemes,
which involve one mechanical and two optical modes, is that the laser
drive is resonant so that, for a fixed drive strength $\Lambda$,
the effective optomechanical coupling strength is much larger. By
contrast, in a three-mode setup, the laser drives are off-resonant
by $\pm\Omega$ so that $\left|\alpha\right|=\frac{\Lambda}{\sqrt{\Omega^{2}+(\kappa/2)^{2}}}$,
whereas in our setup, with $\Delta=0$, we have $\left|\alpha\right|=\frac{2\Lambda}{\kappa}$.
Since the intensity of entangled photons scales with $g^{4}$ and
$g\propto\alpha g_{0}$, this leads to an enhancement of $\sim(\frac{2\Omega}{\kappa})^{4}$
of the photon pair creation rate.

\bibliographystyle{prsty}

\begin{thebibliography}{10}
\bibitem{FOOTNOTEENTANGLEMENT}Note that the phrase ``entanglement
spectrum'' has already a different meaning, which is why we prefer
to use ``spectral density of entanglement''.
\end{thebibliography}

\end{document}